# Grain Boundary Segregation Models for High-Entropy Alloys: Theoretical Formulation and Derived Analytical Expressions to Elucidate High-Entropy Grain Boundaries

*Jian Luo* [1,2,*]

[1] Department of NanoEngineering, University of California San Diego, La Jolla, California 92093, U.S.A.

[2] Program in Materials Science and Engineering, University of California San Diego, La Jolla, California 92093, U.S.A.

**Abstract**

Grain boundary (GB) segregation models are derived for multi-principal element and high-entropy alloys (MPEAs and HEAs). Differing from classical models where one component is taken as solvent and others are considered solutes, these models are referenced to the bulk composition to enable improved treatments of MPEAs and HEAs with no principal components. An ideal solution model is first formulated and solved to obtain analytical expressions that predict GB segregation and GB energy in MPEAs and HEAs. A regular solution model is further derived. The GB composition calculated using the simple analytical expression derived in this study and data from the Materials Project agree well with a prior sphosipcated atomistic simulation for NbMoTaW. The simplicity of the derived analytical expressions makes them useful for not only conveniently predicting GB segregation trends in HEAs, but also analyzing nascent interfacial phenomena in composionally complex GBs. As an application example, the derived models are used to further formulate a set of useful equations to elucidate an emergent concept of high-entropy grain boundaries (HEGBs).

---

[*] Correspondence should be addressed to J.L. at jluo@alum.mit.edu

## I. INTRODUCTION

Grain boundary (GB) segregation (*a.k.a.* adsorption) is one of the most classical and important phenomena in physical metallurgy and ceramics. Various statistical thermodynamic models for GB segregation have been developed and discussed by McLean,[1] Fowler and Guggenheim,[2] Hondros and Seah,[3] Lejček and Hofmann,[4] Wynblatt *et al.*,[5-7] and many others (see, *e.g.*, a book[8] by Lejček). While many of prior models focused on binary systems, statistical thermodynamic models for GB segregation in conventional multicomponent alloys with one principal element, most notably the classical Guttmann model for ternary alloys,[9] have also been derived.

High-entropy alloys (HEAs), which are broadly defined as multi-principal element alloys (MPEAs), have received substantial research interests.[10-13] Understanding GB segregation in HEAs and MPEAs is both scientifically interesting and technologically important. A limited number of atomistic simulations have been performed to model GB segregation in HEAs and MPEAs,[14,15] which generally represent complex and time-consuming tasks. GB segregation in HEAs have also been modeled by simplified density and CALPHAD based approaches.[16,17] Overall, GB segregation in HEAs and MPEAs has not yet been thoroughly investigated, in part due to the compositional complexity that makes quantitative analyses challenging.

Recently, an emergent concept of high-entropy grain boundaries (HEGBs) has been proposed[18] and further elaborated[19] as the GB counterparts to HEAs. It was proposed that effective GB entropy can be positive and increase with increasing number of components in saturated multicomponent alloys (where the chemical potentials are pinned by precipitated secondary phases).[18,19] Such HEGBs can possess reduced GB energy and an increased total adsorption amount with increasing temperature, which can subsequently stabilize nanocystalline alloys (nanoalloys) at high temperatures.[18,19]

Prior statistical thermodynamic models for GB segregation assume one principal component as the solvent with one or more minor component(s) as the solute(s). Differing from classical models, here I formulate GB segregation models referencing to the bulk composition to enable improved treatments of HEAs and MPEAs with no principal component. Such models can also be applied to a conventional multicomponent alloy with one principal component (*e.g.*, to analyze so-called Type I HEGBs in Section V-A). In this work, an ideal solution model is first formulated and solved to obtain analytical expressions to predict GB segregation and GB energy. Subsequently, a regular solution model is derived. It is shown that the GB composition calculated using a simple analytical expression derived in this work and data readily available from the Materials Project agree well with a prior sphosipcated atomistic simulation for NbMoTaW.[20] The models and derived analytical expressions are useful for not only conveniently predicting trends in GB segregation in HEAs, but also analyzing nascent interfacial phenomena in compositionally complex GBs. As an application example, the derived models are used to further formulate a set of useful equations for ideal and symmetric model systems to quantitatively elucidate the emergent concept of HEGBs in Section V.

## II. THEORETICAL FORMULATION

Using the Gibbs approach, GB energy of a multicomponent alloy can be expressed as the interfacial excess of grand potential:

$$\gamma_{GB} = u_{GB}^{XS} - Ts_{GB}^{XS} - \sum_i \Gamma_i \mu_i^{Bulk} \tag{1}$$

The first two terms are the contributions from GB excess of internal energy and GB excess of entropy, respectively. In the third term, $\mu_i$ is the chemical potential and $\Gamma_i$ is the corresponding GB adsorption amount (interfacial excess) of the $i$-th component ($i = 1, 2, 3, \ldots, N$).

The Gibbs adsorption equation states:

$$d\gamma_{GB} = -s_{GB}^{XS} dT - \sum_i \Gamma_i d\mu_i . \tag{2}$$

These are the two most fundamental interfacial thermodynamic equations to analyze GB segregation. It should be noted that GB segregation is interfacial adsorption in thermodynamics. Thus, the terminologies "segregation" and "adsorption" are equivalent and used interchangeably in this article.

## A. Derivation of an Ideal Solution Model and Its Analytical Solution

Let me first consider a large-angle twist GB in a multicomponent ideal solution, as schematically shown in Fig. 1 (similar to a case of a twist GB with $J_{max} = 1$ in the Wynblatt *et al.*'s model for a binary alloy,[7,21] but being generalized for a multicomponent alloy here). The GB excess of internal energy for an ideal solution can be expressed as:

$$u_{GB}^{XS(ideal)} = n \left[ z_v Q \sum_i X_i^{GB} (-e_{ii}) + 2z \sum_i \left( X_i^{GB} - X_i^{Bulk} \right)(-e_{ii}) - 2 \sum_i X_i^{GB} \Delta E_i^{strain} \right]. \tag{3}$$

Here, $n$ is the number of atoms in the atomic plane (so that the number of "GB sites" per unit area is $\Gamma_0 \equiv 2n$ for this twist GB (Fig. 1), assuming the adsorption is limited within the two GB planes, which is largely true for ideal solutions but an approximation for regular solutions), $X_i^{GB}$ and $X_i^{Bulk}$ are the GB and bulk atomic fractions of the $i$-th component, $z$ is the total coordinates (number of bonds per atom), $z_v$ is the number of bonds per atom between two adjacent layers, $e_{ii}$ ($< 0$) is the self-bonding energy of the $i$-th component, $Q$ is the fraction of the broken bonds at the GB core between two GB planes at the twist GB ($Q$ is set to 1/6 to represent an average large-angle general GB to match the empirical relation $\gamma_{GB}/\gamma_{Surface} = 1/3$), and $\Delta E_i^{strain}$ ($> 0$) is the strain energy of the $i$-th component in the bulk.

For a HEA or MPEA without a principal component, $\Delta E_{strain}^i$ should be calculated using the classic Friedel model,[22] but with respect to weighted mean atomic radius and modulus of the bulk phase:

$$\Delta E_i^{strain} = \frac{24\pi K_i \bar{G} \bar{r} r_i (\bar{r} - r_i)^2}{3K_i r_i + 4\bar{G}\bar{r}}, \tag{4}$$

where $r_i$ and $K_i$ are the atomic radius and bulk modulus of the $i$-th component, and $\bar{r}$ and $\bar{G}$ are the weighted means of the atomic radius and shear modulus for the "matrix" phase:

$$\begin{cases} \bar{r} = \sum_k X_k^{Bulk} r_k \\ \bar{G} = \sum_k X_k^{Bulk} G_k \end{cases}. \tag{5}$$

It is noted that this model suggests that the strain energy relaxation can take place for a HEA even without GB segregation. This is reasonable because the lattice distortion can cause significant atomic-scale strains inside a bulk HEA phase even in a random solution.

The GB excess configurational entropy can be expressed as:

$$s_{GB}^{XS} = -\Gamma_0 k \sum_i (X_i^{GB} \ln X_i^{GB} - X_i^{Bulk} \ln X_i^{Bulk}), \tag{6}$$

where $k$ is the Boltzmann constant. The adsorption amount (interfacial excess) of the $i$-th component is given by:

$$\Gamma_i = \Gamma_0 (X_i^{GB} - X_i^{Bulk}). \tag{7}$$

The chemical potential of the $i$-th component in an ideal solution is:

$$\mu_i^{Bulk} = \frac{z}{2} e_{ii} + kT \ln X_i^{Bulk}. \tag{8}$$

Combining the above equations, GB energy for a multicomponent ideal solution can be expressed as:

$$\gamma_{GB} = \Gamma_0 \left[ \tfrac{1}{2} Q z_v \sum_i X_i^{GB} |e_{ii}| - \sum_i X_i^{GB} \Delta E_i^{strain} + \sum_i kT X_i^{GB} \ln \left( \frac{X_i^{GB}}{X_i^{Bulk}} \right) \right]. \tag{9}$$

Using the $j$-th component ($n \neq i$) as a reference, the above equation can be rewritten as:

$$\gamma_{GB} = \gamma_{GB,j}^{(0)} + \Gamma_0 \sum_{i \neq j} X_i^{GB} \left[ \tfrac{1}{2} Q z_v \left( |e_{ii}| - |e_{jj}| \right) - \Delta E_i^{strain} \right] - \Gamma_0 X_j^{GB} \Delta E_j^{strain} + \Gamma_0 \sum_i kT X_i^{GB} \ln \left( \frac{X_i^{GB}}{X_i^{Bulk}} \right), \tag{10}$$

where $\gamma_{GB,j}^{(0)}$ ($= \tfrac{1}{2} \Gamma_0 Q z_v |e_{jj}|$) represents broken-bonds contributed GB energy of the pure component $j$ without any adsorption and relaxation.

Letting $\left( \partial \gamma_{GB} / \partial X_i^{GB} \right)\big|_{X_k^{GB}, k \neq i, j} = 0$ ($dX_i^{GB} = -dX_j^{GB}$ and $dX_k^{GB} = 0$ for all $k \neq i$ or $j$), a Langmuir-McLean type GB adsorption equation [1,3,4] can be derived as:

$$\frac{X_i^{GB}}{X_i^{Bulk}} = \frac{X_j^{GB}}{X_j^{Bulk}} \exp \left( -\frac{\Delta h_{ads.(i \to j)}^{(0)}}{kT} \right). \tag{11}$$

Here, $\Delta h_{ads.(i \to j)}^{(0)}$, where "0" in the subscript denotes the ideal mixing approximation, is defined to represent the enthalpic change associated with swapping one atom of the $i$-th component from the bulk phase with one atom of $j$-th component at the GB:

$$\Delta h^{(0)}_{ads.(i \to j)} = \tfrac{1}{2} Q z_v \left( |e_{ii}| - |e_{jj}| \right) - \left( \Delta E_i^{strain} - \Delta E_j^{strain} \right). \tag{12}$$

Here, a negative segregation enthalpy ($\Delta h^{(0)}_{ads.(i \to j)} < 0$) implies the GB enrichment of the $i$-th component (with respect to the $j$-th component).

Alternatively, the bulk composition can be used as a reference, which can be a better option to treat an HEA (or MPEA) without a principal component. Defining:

$$\overline{e_{self}} = \sum_i X_i^{Bulk} \cdot e_{ii} \tag{13}$$

and

$$\overline{\gamma_{GB}^{(0)}} = \Gamma_0 \sum_i X_i^{Bulk} \cdot \left( \tfrac{1}{2} Q z_v |e_{ii}| \right) = \sum_i X_i^{Bulk} \cdot \gamma_{GB,i}^{(0)}, \tag{14}$$

Equation (9) can be rewritten as:

$$\gamma_{GB} = \overline{\gamma_{GB}^{(0)}} + \Gamma_0 \sum_i X_i^{GB} \left[ \Delta h^{(0)}_{ads.(i)} + kT \ln \left( \frac{X_i^{GB}}{X_i^{Bulk}} \right) \right], \tag{15}$$

where:

$$\Delta h^{(0)}_{ads.(i)} = \tfrac{1}{2} Q z_v \left( |e_{ii}| - |\overline{e_{self}}| \right) - \Delta E_i^{strain}. \tag{16}$$

Differentiation of Equation (15) assuming $dX_i^{GB} = -dX_j^{GB}$ and $dX_k^{GB} = 0$ for $k \neq i, j$ (i.e., swapping one atom of the $i$-th component from the bulk with one atom of $j$-th component at the GB) implies that the following equation holds for any pair of $i$ and $j$ (so that the value of the expression is a constant independent of $i$ or $j$):

$$\Delta h^{(0)}_{ads.(i)} + kT \ln \left( \frac{X_i^{GB}}{X_i^{Bulk}} \right) = \Delta h^{(0)}_{ads.(j)} + kT \ln \left( \frac{X_j^{GB}}{X_j^{Bulk}} \right) = g_0 = kT \ln \kappa_0. \tag{17}$$

Here, $g_0$ and $\kappa_0$ are constants. Thus, $X_i^{GB}$ can be solved as a function of $\kappa_0$:

$$X_i^{GB} = \kappa_0 X_i^{Bulk} e^{-\frac{\Delta h^{(0)}_{ads.(i)}}{kT}}. \tag{18}$$

The constant $\kappa_0$ can be determined by

$$\sum_i X_i^{GB} = \kappa_0 \sum_i X_i^{Bulk} e^{-\frac{\Delta h^{(0)}_{ads.(i)}}{kT}} = 1. \tag{19}$$

Thus, an analytical solution is obtained:

$$X_i^{GB} = \frac{X_i^{Bulk} e^{-\frac{\Delta h_{ads.(i)}^{(0)}}{kT}}}{\sum_i X_i^{Bulk} e^{-\frac{\Delta h_{ads.(i)}^{(0)}}{kT}}}. \tag{20}$$

Combining Equations (15) and (17) produces:

$$\begin{aligned}\gamma_{GB} &= \overline{\gamma_{GB}^{(0)}} + \Gamma_0 \sum_i X_i^{GB} \left[ \Delta h_{ads.(i)}^{(0)} + kT \ln\left(\frac{X_i^{GB}}{X_i^{Bulk}}\right) \right] \\ &= \overline{\gamma_{GB}^{(0)}} + \Gamma_0 \sum_i X_i^{GB} \left( kT \ln \kappa_0 \right) \\ &= \overline{\gamma_{GB}^{(0)}} + \Gamma_0 kT \ln \kappa_0 \end{aligned} \tag{21}$$

Using Equation (19), an analytical solution for GB energy for a multicomponent ideal solution is obtained as follows:

$$\gamma_{GB} = \overline{\gamma_{GB}^{(0)}} - \Gamma_0 kT \ln \left( \sum_i X_i^{Bulk} e^{-\frac{\Delta h_{ads.(i)}^{(0)}}{kT}} \right). \tag{22}$$

## B. Derivation of a Regular-Solution Model

The ideal solution model can be further extended to a regular solution model. For a regular solution, the pair-interaction parameter is defined as:

$$\omega_{ij} = e_{ij} - \tfrac{1}{2}(e_{ii} + e_{jj}). \tag{23}$$

GB energy for a multicomponent regular solution can be expressed as:

$$\gamma_{GB} = \overline{\gamma_{GB}^{(0)}} + \Gamma_0 \sum_i X_i^{GB} \left[ \Delta h_{ads.(i)}^{(0)} + kT \ln\left(\frac{X_i^{GB}}{X_i^{Bulk}}\right) \right] + u_{GB}^{XS(\omega)}, \tag{24}$$

where $u_{GB}^{XS(\omega)}$ is a regular solution (pair-interaction) excess term that should be linearly proportional to the pair-interaction parameters ($\omega_{ij}$'s).

For a regular solution, a Guttmann [23] like adsorption equation can be derived (assuming fixed GB sites of identical adsorption enthalpy for simplicity) as:

$$\frac{X_i^{GB}}{X_i^{Bulk}} = \frac{X_j^{GB}}{X_j^{Bulk}} \exp\left(-\frac{\Delta h_{ads.(i \to j)}}{kT}\right), \tag{25}$$

where:

$$\begin{aligned}
\Delta h_{ads(i \to j)} &= \Delta h^{(\gamma)}_{ads.(i \to j)} + \Delta h^{(\varepsilon)}_{ads.(i \to j)} + \Delta h^{(\omega)}_{ads(i \to j)} \\
&= \Delta h^{(0)}_{ads.(i \to j)} + \Delta h^{(\omega)}_{ads(i \to j)} \\
&= \Delta h^{(0)}_{ads(i)} - \Delta h^{(0)}_{ads(j)} + \Delta h^{(\omega)}_{ads(i \to j)}
\end{aligned} \qquad (26)$$

Here, $\Delta h^{(0)}_{ads.(i \to j)} = \Delta h^{(\gamma)}_{ads.(i \to j)} + \Delta h^{(\varepsilon)}_{ads.(i \to j)}$, expressed in Equation (12) for an ideal solution, includes the contributions from bond/surface energy ($\Delta h^{(\gamma)}_{ads.(i \to j)} = \frac{1}{2} Q z_v \left( |e_{ii}| - |e_{jj}| \right)$) and strain energy ($\Delta h^{(\varepsilon)}_{ads.(i \to j)} = -\left( \Delta E^i_{strain} - \Delta E^j_{strain} \right)$), and it is independent of the composition. The third term (pair-interaction contribution), $\Delta h^{(\omega)}_{ads(i \to j)}$, which will be derived next, depends on the bulk and GB compositions ($X^{Bulk}_k$ and $X^{GB}_k$) and pair-interaction parameters ($\omega_{ij}$'s). Thus, $\Delta h_{ads.(i \to j)}$ is not a constant for a regular solution (*vs.* a constant $\Delta h^{(0)}_{ads.(i \to j)}$ in the Langmuir-McLean type model for an ideal solution), so that Equation (25) represents a Guttmann [23] type adsorption equation (as shown subsequently).

For a regular solution, $\Delta h^{(\omega)}_{ads(i \to j)}$ represents the pair-interaction excess segregation enthalpy (beyond the ideal-solution term) to swap one atom of the *i*-th component in the bulk phase with one atom of *j*-th component at the GB. To assess this term, let me first evaluate the pair-interaction excess internal energy to remove one atom of the *i*-th component in a homogenous bulk phase, which can be expressed as:

$$\Delta u_i^{(\omega-Bulk)} = -z \sum_{k \neq i} X^{Bulk}_k \omega_{ik} . \qquad (27)$$

Here, $\Delta u_i^{(\omega-Bulk)}$ is defined as pair-interaction excess bond energies to create a vacancy via removing one atom of the *i*-th component (by breaking *z* bonds) in a regular solution (where $e_{ij} = \frac{1}{2}(e_{ii} + e_{jj}) + \omega_{ij}$), referenced to (subtracted by) the bond energies in an ideal solution (where $e_{ij} = \frac{1}{2}(e_{ii} + e_{jj})$ and $\omega_{ij} = 0$). Thus, the pair-interaction excess term is $\omega_{ij}$ per *i-j* bond ($e_{ij}$). With this definition, the regular-solution excess enthalpy (internal energy, assuming no change in volume) of mixing per molar can be derived as:

$$H^{XS(regular)}_{mix} = \frac{1}{2} N_A \sum_i X^{Bulk}_i \left( -\Delta u_i^{(\omega-Bulk)} \right) = N_A \frac{z}{2} \sum_i \sum_{k \neq i} X^{Bulk}_i X^{Bulk}_k \omega_{ik} = \sum_{k > i} (z \omega_{ik} N_A) X^{Bulk}_i X^{Bulk}_k = \sum_{k > i} \Omega_{ik} X^{Bulk}_i X^{Bulk}_k , \qquad (28)$$

where $N_A$ is the Avogadro constant and $\Omega_{ik} \equiv z \omega_{ik} N_A$ is the regular-solution parameter. Here, the factor ½ is included because the total bond energies of a bulk phase are twice the sum of broken-bonds energies associated with removing each individual atom (as every bond is shared by two atoms).

Next, the excess pair-interaction energy to remove one atom of the *i*-th component in a region with a compositional gradient (on the *m*-th plane perpendicular to the compositional gradient, without any broken bonds) is given by:

$$\Delta u_i^{(\omega-gradient)} = -\sum_{k \neq i} \left( z_v X_k^{(m-1)} + z_l X_k^{(m)} + z_v X_k^{(m+1)} \right) \omega_{ik}$$
$$= -\left[ z \sum_{k \neq i} X_k^{(m)} \omega_{ik} + z_v \sum_{k \neq i} \left( X_k^{(m-1)} - 2X_k^{(m)} + X_k^{(m+1)} \right) \omega_{ik} \right] \quad (29)$$

where $z_l = z - 2z_v$ is the number of bonds per atom in the lateral atomic plane. After rewriting, the first term in the above equation is $\Delta u_i^{(\omega-Bulk)}$ and the second term is resulted from the compositional gradient. It is further assumed, as a simplification, that GB adsorption is still limited to two GB planes at the core of this twist GB for a regular solution. Taking $X_k^{m-1} = X_k^{Bulk}$ and $X_k^m = X_k^{m+1} = X_k^{GB}$ for a twist GB and considering the existing broken bonds (of fraction $Q$ of $z_v$ bonds per atom) at the twist GB, the excess pair-interaction energy for removing one atom of the $i$-th component at the GB can be deduced as:

$$\Delta u_i^{(\omega-GB)} = -\left\{ \left[ z \sum_{k \neq i} X_k^{GB} \omega_{ik} + z_v \sum_{k \neq i} \left( X_k^{Bulk} - X_k^{GB} \right) \omega_{ik} \right] - Q z_v \sum_{k \neq i} X_k^{GB} \omega_{ik} \right\}$$
$$= -\left\{ \sum_{k \neq i} \left[ z_v X_k^{Bulk} + (z - z_v) X_k^{GB} \right] \omega_{ik} - Q z_v \sum_{k \neq i} X_k^{GB} \omega_{ik} \right\} \quad (30)$$

Subsequently, the pair-interaction excess energy term in the GB segregation enthalpy for swapping one atom of the $i$-th component in the bulk phase with one atom of $j$-th component at the GB can be expressed as the sum of four terms associated with (1) removing one atom of the $i$-th component in the bulk phase ($\Delta u_i^{(\omega-Bulk)}$) to create a vacancy, (2) removing one atom of $j$-th component at the GB ($\Delta u_j^{(\omega-GB)}$) to create a vacancy, (3) inserting one atom of the $i$-th component at the GB to fill the vacancy ($-\Delta u_i^{(\omega-GB)}$), and (4) inserting one atom of $j$-th component in the bulk phase to fill the vacancy ($-\Delta u_j^{(\omega-Bulk)}$), which can be expressed as:

$$\Delta h_{ads(i \to j)}^{(\omega)} = \Delta u_i^{(\omega-Bulk)} + \Delta u_j^{(\omega-GB)} - \Delta u_i^{(\omega-GB)} - \Delta u_j^{(\omega-Bulk)}$$
$$= \left( \Delta u_i^{(\omega-Bulk)} - \Delta u_i^{(\omega-GB)} \right) - \left( \Delta u_j^{(\omega-Bulk)} - \Delta u_j^{(\omega-GB)} \right) \quad (31)$$
$$= \Delta h_{ads(i)}^{(\omega)} - \Delta h_{ads(j)}^{(\omega)}$$

Here, $\Delta h_{ads(i)}^{(\omega)}$ is defined as:

$$\Delta h_{ads(i)}^{(\omega)} \equiv \left( -\bar{u}_i^{(\omega-GB)} \right) - \left( -\bar{u}_i^{(\omega-Bulk)} \right)$$
$$= (z - z_v) \sum_{k \neq i} (X_k^{GB} - X_k^{Bulk}) \omega_{ik} - Q z_v \sum_{k \neq i} X_k^{GB} \omega_{ik} \quad (32)$$
$$= \left[ z - (1+Q) z_v \right] \sum_{k \neq i} X_k^{GB} \omega_{ik} - (z - z_v) \sum_{k \neq i} X_k^{Bulk} \omega_{ik}$$

It represents the difference in the pair-interaction excess energy terms for removing one atom of $i$-th component in the bulk phase *vs.* removing one atom of $i$-th component at the GB. Thus, the pair-interaction excess GB segregation enthalpy can be expressed as:

$$\begin{aligned}\Delta h^{(\omega)}_{ads(i\to j)} &= \Delta h^{(\omega)}_{ads(i)} - \Delta h^{(\omega)}_{ads(j)} \\ &= \left[(z-z_v)\sum_{k\ne i}(X_k^{GB}-X_k^{Bulk})\omega_{ik} - Qz_v\sum_{k\ne i}X_k^{GB}\omega_{ik}\right] - \left[(z-z_v)\sum_{k\ne j}(X_k^{GB}-X_k^{Bulk})\omega_{jk} - Qz_v\sum_{k\ne j}X_k^{GB}\omega_{ik}\right] \\ &= \left[(z-z_v)\sum_{k\ne i}(X_k^{GB}-X_k^{Bulk})\omega_{ik} - Qz_v X_j^{GB}\omega_{ij}\right] - \left[(z-z_v)\sum_{k\ne j}(X_k^{GB}-X_k^{Bulk})\omega_{jk} - Qz_v X_i^{GB}\omega_{ij}\right] \\ &= \left[(z-z_v)\sum_{k\ne i}(X_k^{GB}-X_k^{Bulk})\omega_{ik} + Qz_v X_i^{GB}\omega_{ij}\right] - \left[(z-z_v)\sum_{k\ne j}(X_k^{GB}-X_k^{Bulk})\omega_{jk} + Qz_v X_j^{GB}\omega_{ij}\right]\end{aligned} \quad (33)$$

Appendix II shows that a simplified case of this derived expression for $N = 2$ is fully consistent with the Wynblatt *et al.*'s model for binary alloys.[7,21] Thus, the current model can be considered as a multicomponent generalization of the Wynblatt *et al.*'s binary alloy model for a case of bilayer adsorption at a twist GB.[7,21] Using $X_j^{GB} = 1 - \sum_{k\ne j} X_k^{GB}$, the above equation can also be rewritten as:

$$\Delta h^{(\omega)}_{ads(i\to j)} = (z-z_v)\left[-2(X_i^{GB}-X_i^{Bulk})\omega_{ij} + \sum_{k\ne i,j}(X_k^{GB}-X_k^{Bulk})(\omega_{ik}-\omega_{jk}-\omega_{ji})\right] + Qz_v\left(X_i^{GB}-X_j^{GB}\right)\omega_{ij} \quad (34)$$

Here, the first term in the above equation is equivalent to that in Equation (5b) in Ref. 25 (by taking $X_i^{m-1} = X_i^{Bulk}$ and $X_i^m = X_i^{m+1} = X_i^{GB}$; noting "$k$", instead of "$m$", was used, and $j = 1$, in Equation (5b) in Ref. 25). Note that a broken-bonds contribution term is also included in the above equation (but not in Equation (5b) in Ref. 25).

The above equation can be rewritten to a form similar to the Guttmann model,[23] as:

$$\Delta h_{ads(i\to j)} = \Delta \tilde{h}^{(0)}_{ads(i\to j)} - 2\hat{\alpha}_{ij} X_i^{GB} + \sum_{k\ne i,j} \hat{\alpha}'_{ik-j} X_k^{GB}. \quad (35)$$

However, $\hat{\alpha}'_{ik-j} \ne \hat{\alpha}_{ik} - (\hat{\alpha}_{ij} + \hat{\alpha}_{kj})$ because of the extra broken-bonds term (*i.e.*, the GB is no longer a regular solution with the broken bonds, so the above equation does not follow the regular solution based Guttmann model[23] exactly).

For a regular solution, GB segregation (*a.k.a.* adsorption) enthalpy can be defined as:

$$\Delta h_{ads.(i)} \equiv \Delta h^{(0)}_{ads.(i)} + \Delta h^{(\omega)}_{ads.(i)}. \quad (36)$$

Similar to Equation (17) and (20), the following two equations can also be derived:

$$\Delta h_{ads.(i)} + kT\ln\left(\frac{X_i^{GB}}{X_i^{Bulk}}\right) = \Delta h^{(0)}_{ads.(i)} + \Delta h^{(\omega)}_{ads.(i)} + kT\ln\left(\frac{X_i^{GB}}{X_i^{Bulk}}\right) = \text{constant}, \quad (37)$$

and

$$X_i^{GB} = \frac{X_i^{Bulk} e^{-\frac{\Delta h_{ads.(i)}}{kT}}}{\sum_i X_i^{Bulk} e^{-\frac{\Delta h_{ads.(i)}}{kT}}} = \frac{X_i^{Bulk} e^{-\frac{\Delta h^{(0)}_{ads.(i)}+\Delta h^{(\omega)}_{ads.(i)}}{kT}}}{\sum_i X_i^{Bulk} e^{-\frac{\Delta h^{(0)}_{ads.(i)}+\Delta h^{(\omega)}_{ads.(i)}}{kT}}}. \quad (38)$$

For a regular solution, the GB composition ($X_i^{GB}$) needs to be solved numerically using the above equation because $\Delta h_{ads(i)}^{(\omega)}$ also depends on $X_i^{GB}$. If the pair-interaction term $\Delta h_{ads(i)}^{(\omega)}$ is small, the above equation can be solved iteratively. Otherwise, one should search for the initial $X_i^{GB}$ values that lead to convergence of iterating Equation (38) or use a better algorithm to solve Equation (38) numerically.

Practically, using the regular solution model needs to quantify $N(N-1)/2$ pair-interaction parameters and solving Equation (38) numerically, which may not be trivial for a HEA. Thus, the analytical expression shown in Equation (22) for an ideal solution, which can be readily quantified and produce results agree well with the atomistic simulations for a HEA (as shown in the next section), can be taken as the first-order approximation to predict useful trends.

## C. Further Generalization

More rigorously, segregation entropy (vibration and other contributions, beyond the configurational entropy that is already considered) is also important, as discussed by Lejček *et al.*,[26-28] and non-regular interaction terms can also exist. Thus, Equations (11), (20), (25) and (38) can be further generalized to:

$$\frac{X_i^{GB}}{X_i^{Bulk}} = \frac{X_j^{GB}}{X_j^{Bulk}} \exp\left(-\frac{\Delta g_{ads.(i \to j)}}{kT}\right), \tag{39}$$

and

$$X_i^{GB} = \frac{X_i^{Bulk} e^{-\frac{\Delta g_{ads.(i)}}{kT}}}{\sum_i X_i^{Bulk} e^{-\frac{\Delta g_{ads.(i)}}{kT}}}. \tag{40}$$

Here, $\Delta g_{ads.(i \to j)} = \Delta h_{ads.(i \to j)} - T \Delta s_{ads.(i \to j)}^{excess}$, where $\Delta h_{ads.(i \to j)}$ is the segregation enthalpy (that can also include enthalpically interaction terms beyond the regular solution model) and $\Delta s_{ads.(i \to j)}^{excess}$ is the segregation entropy beyond the already-considered configurational entropy (*i.e.*, contributions from phonons/vibration, electron, magnetism, *etc.*).

The following equation for GB energy, which has been rigorously proved for ideal solutions, may only be used as an approximation when $\Delta g_{ads.(i \to j)}$ is not a constant:

$$\begin{aligned}
\gamma_{GB} &\approx \overline{\gamma_{GB}^{(0)}} + \Gamma_0 \sum_i X_i^{GB} \left[ \Delta g_{ads.(i)} + kT \ln\left(\frac{X_i^{GB}}{X_i^{Bulk}}\right) \right] \\
&\approx \overline{\gamma_{GB}^{(0)}} - \Gamma_0 kT \ln\left( \sum_i X_i^{Bulk} e^{-\frac{\Delta g_{ads.(i)}}{kT}} \right)
\end{aligned} \tag{41}$$

### III. A CALCULATION EXAMPLE

Using NbMoTaW as an example, GB segregation was computed using the analytical expressions derived for an ideal solution (Equations (20) and (22)) and the density functional theory (DFT) data of the lattice parameters, bond energies, bulk and shear moduli taken from the Materials Project.[29] The data and calculated relevant thermodynamic quantities are shown in Table 1 for an equimolar NbMoTaW. Figure 2 shows the computed GB composition and GB energy as functions of temperature for this equimolar NbMoTaW. Strong segregation of Nb and depletion of W and Ta, particularly at low temperatures, were predicted for the equimolar NbMoTaW by this model (Fig. 2). For the fixed GB composition, the lattice model predicted that GB energy increases with increasing temperature due to temperature-induced desorption (Fig. 2b).

In general, the adoption of readily available data from the Materials Project allows consistent and convenient parameterization of the model. Along with the simple analytical expressions derived in this study, segregation trends can be forecasted for HEAs and MPEAs with ease, which are useful.

To further benchmark the model, the calculated results from the simple analytical solution derived in this work was compared with a prior hybrid Monte Carlo and molecular dynamics (MC/MD) simulation of a polycrystal with six randomly inserted GBs at 300 K conducted by Li *et al.*[20] in Table 3. In the hybrid MC/MD simulation of the polycrystal, the bulk (grain) composition deviated from the equimolar composition due to GB segregation and a small average grain diameter of ~7.5 nm in a closed system (even if the overall composition is equimolar in the 11 nm × 11 nm × 11 nm simulation supercell). For example, the predicted GB segregation of Nb, which was indeed observed in the hybrid MC/MD simulation, resulted in the depletion of Nb inside the grains in the hybrid MC/MD simulation. To do a fair comparison, the analytical expression was used to calculate GB composition ($X_i^{GB}$) for the non-equimolar bulk composition of $Nb_{0.155}Mo_{0.246}Ta_{0.280}W_{0.319}$, which matched the observed average grain composition in the MC/MD simulation. The (110) twist GB ($z_v = 2$ and $Q = 1/6$) was selected to represent a general large-angle GB with moderate segregation ($z_v / z = ¼$). Note that the weighted average radius, modulus, and bond energy, as well as strain energies and segregation enthalpies calculated for this non-equimolar alloy (Table 2) are slightly different from those calculated for the equimolar NbMoTaW (Table 1).

As shown in Table 3, the calculated GB composition (~68% Nb, ~24% Mo, ~6% Ta, and ~1% W) using a simple analytical expression derived in this study agrees well with the GB composition (~58% Nb, ~27% Mo, ~14% Ta, and ~2% W) obtained by the sophisticated hybrid MC/MD simulation.[20] Not only the relative segregation trend is correct, the calculated GB composition using the analytical expression (Equations (20)), even if it is derived from a simplified ideal solution model, quantitatively agrees with the sophisticated hybrid MC/MD simulation (that needed first fitting a DFT supported machine learning potential)[20] within reasonable errors.

This success suggests that the derived analytical expressions, using the data readily available from the Materials Project,[29] can be robustly useful to predict trends of GB segregation in HEAs and MPEAs.

### IV. DISCUSSION OF THE DERIVED MODELS

GB segregation in HEAs have been so far rarely characterized or modelled because of the compositional complexity. The derived analytical expressions for the simplified ideal solution model, along with readily available high-quality data from the Materials Project,[29] present a facile method to quickly forecast GB segregation trends in HEAs and MPEAs. This model can be further extended to predict GB segregation in other types of GBs (beyond the twist GBs analyzed here), as well as anisotropic segregation (after deriving the relevant GB segregation enthalpy expressions for multicomponent systems in a follow-up study). The model for regular solutions has also been derived. However, obtaining consistent and reliable data and developing robust numerical algorithms to solve the regular solution model for HEAs can still be challenging. This should be investigated further in future studies. In this regard, the derived analytical expressions for the simplified ideal solution model, which can be readily parameterized by the high-quality data from the Materials Project,[29] is robustly useful. The predictability has been shown for NbMoTaW.

The current models also have a few limitations. The analytical solutions are limited to the case of fixed number of GB adsorption sites with an identical segregation enthalpy. It is known that multilayer adsorption, layering and prewetting transitions, and GB critical phenomena, can take place in strong segregation region in regular solutions.[6,21,24,30,31] Such interesting interfacial phenomena can in principle be solved numerically with the lattice models (as shown for simpler dilute binary solutions with a principal component).[6,21,24,30,31] However, it is still challenging to analyze such complex interfacial phenomena in HEAs and MPEAs due to the compositional complexity. Moreover, the lattice models do not consider interfacial structural transitions,[32-46] including GB disordering (particularly the possible formation of liquid-like GBs in coupled premelting and prewetting regions),[34-36,44,46-51] which can take place and affect GB adsorption and GB energy. Future studies should be conducted to consider and investigate these effects, which are highly challenging to model together with the compositional complexity. Yet, the simplicity of the analytical expressions (particularly Equations (15), (20), and (22)) derived in this study for simplified models makes them robustly useful.

In addition to predicting useful trends for GB segregation in HEAs and MPEAs via analytical expressions, the derived models and expressions can also be used to analyze nuisance interfacial phenomena in compositionally complex GBs. As an application example, the derived analytical expressions are used to further formulate a set of simplified equations to elucidate the emergent concept of HEGBs subsequently.

## V. APPLICATION TO ELUCIDATE HEGBS

The rapid development of HEAs simulated a scientific interest to explore the existence and character of their interfacial (GB) counterparts – "high-entropy grain boundaries (HEGBs)". The concept of HEGBs was originally proposed in a research article[18] in 2016, with a goal to utilize such HEGBs to stabilize nanoalloys against grain growth at high temperatures (Fig. 3a). The relevant concepts were further elaborated in a short perspective article in 2023.[19] However, rigorous full derivations have not been given in detail, in part due to the lacking of a useful GB segregation model for analyzing multicomponent alloys and HEAs. Here, such a model has been derived in this study. Thus, the analytical expressions derived here are used to further formulate equations to elucidate the emergent concept of HEGBs, with rigorous full derivations.

First, what is GB entropy? This may not be a trivial question. There are two general ways to define GB entropy.

In the Gibbs approach, GB excess of entropy ($s_{GB}^{XS}$) is a well-defined quantity independent of the selection of the Gibbs dividing plane (differing from surfaces and other heterointerfaces with two different abutting phases). GB excess of entropy ($s_{GB}^{XS}$) appears in the Gibbs adsorption equation (Equation (2)), a fundamental interfacial thermodynamic relation governing GB segregation (*a.k.a.* interfacial adsorption).

In an alternative and perhaps more useful definition analogous the bulk thermodynamic relation $dG = -SdT + VdP$ (so that $S = -(\partial G/\partial T)_P$), one can define effective GB entropy to characterize temperature dependence of GB energy:

$$s_{GB}^{\text{eff.}} \equiv -\left(\frac{\partial \gamma_{GB}}{\partial T}\right)_{P,\ldots}. \tag{42}$$

The effective GB entropy for a unary GB can be derived from Equation (2), as:[19]

$$s_{GB}^{\text{eff.}} = -\left(\frac{\partial \gamma_{GB}}{\partial T}\right)_P = s_{GB}^{XS} + \Gamma\left(\frac{\partial \mu}{\partial T}\right)_P = s_{GB}^{XS(\text{disorder})} + S_V \cdot \Delta v_{GB}^{free}, \tag{43}$$

where $S_V$ is the volumetric entropy (*i.e.*, the molar entropy $\overline{S}$ divided by the molar volume $\overline{V}$) and $\Delta v_{GB}^{free}$ is the GB free volume ($\Delta v_{GB}^{free} = \Gamma \cdot \overline{V}$). For a unary GB, $s_{GB}^{XS}$ is mostly from GB disorder (with structural, vibration, electronic, and other contributions), so it is denoted as $s_{GB}^{XS(\text{disorder})}$. This equation implies that the effective GB entropy is not equal to GB excess of entropy ($s_{GB}^{\text{eff.}} \neq s_{GB}^{XS}$) even for a simple unary GB. For a unary system, the effective GB entropy is usually positive due to the interfacial disordering and GB free volume, as shown in the above equation. However, these structural effects are not considered in the lattice models (with no interfacial disordering and GB free volume). Thus, additional effective GB entropy terms similar to these two terms in the above equation should be added separately to the lattice models (discussed below) for a multicomponent GB:

$$s_{GB}^{\text{eff.}} = -\left(\frac{\partial \gamma_{GB}}{\partial T}\right)_{P, X_i \text{ or } \mu_i} \approx s_{GB}^{\text{eff.(ads.)}} + s_{GB}^{XS(\text{disorder})} + \sum_i \Gamma_i \cdot \overline{S}_i + \text{coupling terms}, \tag{44}$$

where $s_{GB}^{\text{eff.(ads.)}}$ is a term representing the GB segregation effects that will be discussed in detail subsequently. Similar to a unary GB, the second term is resulted from GB disorder (including structural as well as vibration, electronic, and other, e.g., magnetic, contributions) and the third term is correlated to GB free volume ($\Delta v_{GB}^{free} = \sum_i \Gamma_i \cdot \overline{V}_i$ for a multicomponent GB). In addition, coupling terms, *e.g.*, that between GB segregation and GB disordering, can generally exist and often be significant (particularly with the possible formation of liquid-like GBs at high temperatures[34-36,44,46-51]).

The following discussion focuses on the adsorption (GB segregation) effects on the effective GB entropy (or the $s_{GB}^{\text{eff.(ads.)}}$ term in the above equation). Equation (2) suggests that GB segregation (*a.k.a.* interfacial adsorption in thermodynamics) reduces GB energy. This effect is well known and has been utilized to stabilize nanoalloys.[52-62] However, temperature-induced GB desorption can counter this stabilization effect (Fig. 3b) at high temperatures. Thus, it is beneficial to seek for conditions where GB energy reduces and GB segregation increases with increasing temperature in so-called HEGBs, which can offer both reduced thermodynamic driving pressures and increased kinetic solute-drag[63] pressures to stabilize nanoalloys (Fig. 3b).[19] In addition, a unique feature of HEGBs should be that such effects (*e.g.*, positive effective GB entropy) should increase with increasing number of components.

For a single-phase multicomponent alloy with a fixed bulk composition and a fixed number of GB adsorption sites, GB energy increases with increasing temperature due to temperature-induced desorption or de-segregation (Fig. 4a). This effect is seen in Fig. 2(b) for NbMoTaW computed from the lattice model (albeit that effects of interfacial disordering and GB free volume, which are not considered in the lattice model, can reduce GB energy with increasing temperature). Thus, the effective GB entropy (from the adsorption contribution in the lattice model) is negative:[19]

$$s_{GB}^{eff.\,(ads.,\,\text{fixed}\,X_i^{Bulk})} = -\left(\frac{\partial \gamma_{GB}}{\partial T}\right)_{P, X_i^{Bulk}} < 0. \tag{45}$$

HEGBs cannot exist from GB adsorption effects in this scenario. However, the formation of HEGBs can be promoted in two scenarios, as shown in Fig. 4b and 4c and explained below.

On the one hand, interfacial disordering (and widening), particularly the formation of liquid-like GBs in the coupled premelting and prewetting regions,[34-36,44,46-51] may enhance GB adsorption (Fig. 4b). Interfacial disordering and GB free volume is known to decrease GB energy with increasing temperature to produce a positive effective GB entropy in unary GBs based on Equation (43) and a prior atomistic simulation of Ni GBs.[64] Such effects should also exist in multicomponent GBs, but are not considered in the lattice models. Moreover, it is possible that the coupling of interfacial disordering and adsorption can promote and enhance the formation of HEGBs, particularly with the formation of liquid-like GBs in the coupled premelting and prewetting regions.[34-36,44,46-51] However, this hypothesis has yet been directly confirmed or rigorously approved.

On the other hand, it was proposed[18] and recently elaborated a perspective article[19] that GB adsorption alone can lead to the formation of HEGBs in saturated alloys (Fig. 4c) with chemical potentials pinned (fixed) by precipitated secondary phases to produce positive effective GB entropy:[19]

$$s_{GB}^{eff.\,(ads.,\,\text{saturated})} = -\left(\frac{\partial \gamma_{GB}}{\partial T}\right)_{P, X_i^{Bulk}\,\text{on the max solvus}} \approx -\left(\frac{\partial \gamma_{GB}}{\partial T}\right)_{P, \mu_i} > 0. \tag{46}$$

Such HEGBs may possess three characters:

(1) GB energy reduces with increasing temperature,

(2) the total GB adsorption amount increases with increasing temperature, and

(3) the positive effective GB entropy increases with increasing number of components.

Two basic types of HEGBs are envisioned.[19] Type I HEGBs can form in a saturated conventional multicomponent alloy with $(N-1)$ precipitated secondary phases that pin the chemical potentials of the primary phase (Fig. 4c). Type II HEGBs can form in a HEA or MPEA with $(N-1)$ principal elements and one strongly segregating element, where one or more precipitated secondary phase(s) pins the bulk chemical potential(s).

Subsequently, the newly derived analytical expressions are used to elucidate Type I and Type II HEGBs, as well as the general mixed cases, with full derivations that had not been provided previously.[19]

## A. Analysis of Type I HEGBs

Let me first consider a conventional multicomponent alloy with one principal element (Component 1 as the solvent) and $(N-1)$ segregating solute elements. Based on Equation (22) (or (41) as a generalization, rigorously held for an ideal solution) and using the relation $\Delta g_{ads.(i \to 1)} \equiv \Delta g_{ads.(i)} - \Delta g_{ads.(1)}$, the following equation can be obtained:

$$\begin{aligned}
\gamma_{GB} &= \overline{\gamma_{GB}^{(0)}} - \Gamma_0 kT \ln \left( X_1^{Bulk} e^{-\frac{\Delta g_{ads.(1)}}{kT}} + \sum_{i=2}^{N} X_i^{Bulk} e^{-\frac{\Delta g_{ads.(i)}}{kT}} \right) \\
&= \overline{\gamma_{GB}^{(0)}} - \Gamma_0 kT \ln \left[ e^{-\frac{\Delta g_{ads.(1)}}{kT}} \left( X_1^{Bulk} + \sum_{i=2}^{N} X_i^{Bulk} e^{-\frac{\Delta g_{ads.(i)}}{kT}} e^{+\frac{\Delta g_{ads.(1)}}{kT}} \right) \right] \\
&= \overline{\gamma_{GB}^{(0)}} + \Gamma_0 \Delta g_{ads.(1)} - \Gamma_0 kT \ln \left( X_1^{Bulk} + \sum_{i=2}^{N} X_i^{Bulk} e^{-\frac{\Delta g_{ads.(i \to 1)}}{kT}} \right) \\
&\approx \overline{\gamma_{GB}^{(0)}} - \Gamma_0 kT \ln \left( 1 + \sum_{i=2}^{N} X_i^{Bulk} e^{-\frac{\Delta g_{ads.(i \to 1)}}{kT}} \right) \qquad (\text{as } X_1^{Bulk} \to 1)
\end{aligned} \qquad (47)$$

Here, $\Delta g_{ads.(1)}$ vanishes as $X_1^{Bulk} \to 1$ in the bulk dilute solution limit.

Considering a case of fixed bulk composition with $(N-1)$ segregating solute elements of identical $X_i^{Bulk}$ and $\Delta g_{ads.(i \to 1)}$, the reduction of GB energy upon segregation depends on the total amount of solute atoms $\sum_{i=2}^{N} X_i^{Bulk}$, but not on the number of solute components $(N-1)$. Thus, there is no HEGB effect for the GB adsorption contribution in this fixed bulk composition scenario.

However, HEGBs can exist in a fixed chemical potentials scenario, represented by a saturated multicomponent alloy.[19] In this scenario of Type I HEGBs, the chemical potentials are pinned by $(N-1)$ precipitated secondary phases in an *N*-component system (with no compositional degree of freedom according to the Gibbs phase rule), so that the bulk composition moves along the maximum solvus line with increasing temperature. Assuming a precipitate of a binary $M_xS_y$ line compound (M = matrix component 1 and S = solute component *i*), the solvus for a dilute solution ($X_1 \to 1$) in the 1-*i* binary system can be expressed (based on the derivations in Appendix III) as:

$$X_{i(1)}^{Binary\ Solvus} \approx \alpha_{i(1)}^{Binary} \left( X_1 \right)^{-\beta_{i(1)}} e^{\frac{\Delta g_{i(1)}^{ppt.}}{kT}} \approx \alpha_{i(1)}^{Binary} e^{\frac{\Delta g_{i(1)}^{ppt.}}{kT}}, \qquad (48)$$

where is $\Delta g_{i(1)}^{ppt.}$ is the free energy of precipitation per atom of the *i*-th component. Here, $\alpha_{i(1)}^{Binary}$ is a constant for a dilute solution obeying Henry's law, and $\alpha_{i(1)}^{Binary} = 1$ for an ideal solution. Assuming a dilute ideal multicomponent solution (where $X_i^{Bulk} \approx X_{1-i}^{Binary\ Solvus} = \exp[\Delta g_{i(1)}^{ppt.} / (kT)]$, *i.e.*, the solubility of the *i*-th component in the multicomponent alloy can be approximated by its solubility in the 1-*i* binary system), plugging Equation (48) into Equation (47) produces:

$$\gamma_{GB} = \overline{\gamma_{GB}^{(0)}} - \Gamma_0 kT \ln\left(1 + \sum_{i=2}^{N} e^{\frac{\Delta g_{i(1)}^{ppt.}}{kT}} e^{-\frac{\Delta g_{ads.(i \to 1)}}{kT}}\right)$$

$$= \overline{\gamma_{GB}^{(0)}} - \Gamma_0 kT \ln\left(1 + \sum_{i=2}^{N} e^{-\frac{\Delta g_{i(1)}^{ads.-ppt.}}{kT}}\right) \tag{49}$$

where $\Delta g_{i(1)}^{ads.-ppt.} \equiv \Delta g_{ads.(i \to 1)} - \Delta g_{i(1)}^{ppt.}$ (taking to be a positive value here) represents the free energy to swap one atom of the *i*-th component at the GB with one atom of the 1st component in the precipitate. A more rigorous solution for a regular solution is:

$$\gamma_{GB} = \overline{\gamma_{GB}^{(0)}} - \Gamma_0 kT \ln\left(1 + \sum_{i=2}^{N} \phi_i e^{-\frac{\Delta g_{i(1)}^{ads.-ppt.}}{kT}}\right), \tag{50}$$

where $\phi_i$ ($\to$ 1 for a dilute ideal solution) is a dimensionless coefficient:

$$\phi_i = \left(\frac{X_i^{Bulk}}{X_{i(1)}^{Binary\ Solvus}}\right) \alpha_{i(1)}^{Binary} \left(X_1\right)^{-\beta_{i(1)}}. \tag{51}$$

Here, $X_i^{Bulk} / X_{i(1)}^{Binary\ Solvus}$ (= 1 for an ideal solution) is a ratio representing the co-doping influence on the solid solubility of the *i*-th component. In addition, $\alpha_{i(1)}^{Binary} = 1$ for an ideal solution and $\left(X_1\right)^{-\beta_{i(1)}}$ approaches to unity for a dilute solution.

An empirical relationship exist for most binary alloys:[65]

$$\Delta g_{i(j)}^{ads.-ppt.} \approx \Delta h_{i(j)}^{ads.-ppt.} \approx 0.10 \pm 0.06 \text{ eV/atom} \left(\approx 10 \pm 6 \text{ kJ/mol}\right), \tag{52}$$

Note that a negative sign was used for $\Delta g_{i(j)}^{ads.-ppt.}$ or $\Delta h_{i(j)}^{ads.-ppt.}$ in prior publications,[19,65,66] but it should be positive with this definition.

For Type I HEGBs, assumed $\phi_i \approx 1$ and all $\Delta g_{i(1)}^{ads.-ppt.} = \Delta g_{S(1)}^{ads.-ppt.}$ is a constant for all $i \neq 1$ for simplicity, the following expressions can be obtained to illustrate the characteristics of Type I HEGBs (*i.e.*, reduced GB energy with increasing temperature and increasing number of components):

$$\gamma_{GB} \approx \overline{\gamma_{GB}^{(0)}} - \Gamma_0 kT \ln\left[1 + (N-1) e^{-\frac{\Delta g_{S(1)}^{ads.-ppt.}}{kT}}\right]. \tag{53}$$

Here, $\Delta g_{S(1)}^{ads.-ppt.}$ (where "S" in the subscript represents Solute $i = 2, \ldots, N$) can be taken to be the average value for the segregating species, which should be $\sim 0.10 \pm 0.06$ eV/atom. Thus,

$$s_{GB}^{eff.\,(ads.,\,\text{saturated})} = -\left(\frac{d\gamma_{GB}}{dT}\right) \approx \Gamma_0 \left\{ k \ln\left[ f_I(\Delta g_{S(1)}^{ads.-ppt.}, N-1) \right] + \frac{\Delta g_{S(1)}^{ads.-ppt.}}{T}\left[ 1 - f_I(\Delta g_{S(1)}^{ads.-ppt.}, N-1)^{-1} \right] \right\} > 0, \quad (54)$$

where $f_I(\Delta g_{S(1)}^{ads.-ppt.}, N-1) \equiv 1 + (N-1)\exp[-\Delta g_{S(1)}^{ads.-ppt.}/(kT)]$ (so that $\gamma_{GB} \approx \overline{\gamma_{GB}^{(0)}} - \Gamma_0 kT \ln f_I$) is a dimensionless factor that increases with increasing $(N-1)$. Thus, $s_{GB}^{eff.\,(ads.,\,\text{saturated})}$ increases with increasing number of segregating components.

Fig. 5 shows an analysis of Type I HEGBs based on Equation (53) for a hypothetic saturated multicomponent alloy with one principal component and $(N − 1)$ segregating minor components, where $(N − 1)$ precipitated secondary phases pin the chemical potentials. Specifically, computed GB energy reduction vs. temperature curves for $\Delta g_{S(1)}^{ads.-ppt.} = 0.1$ eV/atom, 0.05 eV/atom, and 0.15 eV/atom, respectively, are shown in Fig. 5. In all cases, GB energy decreases with increasing temperature (producing a positive effective GB entropy, $-d\gamma_{GB}/dT$) in the saturated multicomponent alloy. Furthermore, the effective GB entropy (the negative slope of the curve) increases with increasing number of components. In Fig. 5, the reductions in GB energies are plotted in normalized parameter $(\gamma_{GB} - \overline{\gamma_{GB}^{(0)}})/\Gamma_0$ to represent more general results. For a reference value, if the $\Gamma_0$ value for the Ni (100) twist GB is adopted, $(\gamma_{GB} - \overline{\gamma_{GB}^{(0)}})/\Gamma_0 = -0.18$ eV corresponds to a $(\overline{\gamma_{GB}^{(0)}} - \gamma_{GB})$ value of ~0.969 J/m², which represents a substantial reduction in GB energy.

An isothermal section of ternary phase diagram displaying the saturated composition for binary ($N − 1 = 1$) and ternary ($N − 1 = 2$) alloys are also shown in Fig. 5. This ternary phase diagram illustrates the increased total solid solubility of all solutes in the primary bulk phase from a binary alloy (represented by the black dots) to a ternary alloy (represented by the red dot), which results in the increased total GB adsorption (segregation amount) to reduce GB energy. The same mechanism also exists and is expected to provide HEGB effects for $N > 3$.

The general characters of Type I HEGBs are well illustrated in Fig. 5 for the simplified cases of identical binary subsystems in normalized parameters. For Type I HEGBs, Equations (49) and (50) can be applied to forecast multicomponent alloys with different $\Delta g_{i(1)}^{ads.-ppt.}$ values for different segregating solutes and precipitates. More realistic modeling of real multicomponent alloys and Type I HEGBs can be conducted by applying Equation (22) directly using the maximum solvus composition as the bulk composition, which can be obtained by bulk CALPHAD methods. If the multicomponent maximum solvus composition cannot be determined (due to lacking CALPHAD data), the following approximation (that is rigorously held for ideal solutions) can be adopted as a first-order approximation.

$$X_i^{Bulk} \approx X_{i(1)}^{Binary\ Solvus}. \quad (55)$$

It should be noted that in this case that $X_i^{Bulk}$ ($i \neq 1$) is different from (smaller than) the overall (initial)

composition due to the precipitation. This approach enables us to use an ideal solution with the bonding energies and segregation enthalpies mimicking the real multicomponent alloy, along with real binary phase diagrams, to predict useful trends.

## B. Analysis of Type II HEGBs

Next, let me consider a HEA or MPEA with $(N-1)$ principal elements (Components 1, 2, … $N-1$) and one strongly segregating solute element (Component $N$). For this case, Equation (22) or its approximated generalized version Equation (41) (rigorously held only for ideal solutions) can be rewritten as:

$$\begin{aligned}
\gamma_{GB} &= \overline{\gamma_{GB}^{(0)}} - \Gamma_0 kT \ln\left( \sum_{i=1}^{N-1} X_i^{Bulk} e^{-\frac{\Delta g_{ads.(i)}}{kT}} + X_N^{Bulk} e^{-\frac{\Delta g_{ads.(N)}}{kT}} \right) \\
&= \overline{\gamma_{GB}^{(0)}} - \Gamma_0 kT \ln\left[ e^{-\frac{\Delta g_{ads.(N)}}{kT}} \left( \sum_{i=1}^{N-1} X_i^{Bulk} e^{-\frac{\Delta g_{ads.(i)}}{kT}} e^{+\frac{\Delta g_{ads.(N)}}{kT}} + X_N^{Bulk} \right) \right] \\
&= \overline{\gamma_{GB}^{(0)}} + \Gamma_0 \Delta g_{ads.(N)} - \Gamma_0 kT \ln\left( \sum_{i=1}^{N-1} X_i^{Bulk} e^{+\frac{\Delta g_{ads.(N \to i)}}{kT}} + X_N^{Bulk} \right)
\end{aligned} \qquad (56)$$

Here, $\Delta g_{ads.(N)}$ is not small (and cannot be neglected). If $X_N^{Bulk} \ll 1$. If only the $N$-th component is strongly segregating (*i.e.*, $\Delta g_{ads.(i)}$ is negligibly small for all principal elements so that $\Delta g_{ads.(N)} \approx \Delta g_{ads.(N \to i)}$) and $X_N^{Bulk} \ll 1$ (so that $\sum_{i=1}^{N-1} X_i^{Bulk} \approx 1$), the following approximation can be derived:

$$\gamma_{GB} \approx \overline{\gamma_{GB}^{(0)}} - \Gamma_0 kT \ln\left( 1 + X_N^{Bulk} e^{-\frac{\Delta g_{ads.(N \to i)}}{kT}} \right). \qquad (57)$$

Unlike the case of Type I HEGBs discussed previously, the solubility of the $N$-th component in the HEA, $X_N^{Bulk}$, is significantly greater than the binary solid solubility.

Similar to Type I HEGBs, bulk high-entropy effects can exist for Type II HEGBs in a fixed chemical potential scenario to reduce GB energy with increasing temperature, represented by saturated HEAs.[19,66]

For a selected HEA composition (assuming that $X_N^{Bulk} \ll 1$ so that the precipitation does not change the bulk composition of principal components significantly), the chemical potential of the $N$-th component will be pinned by the precipitation of the first line compound (Appendix III). Thus, the solubility of the $N$-th component in the HEA (or MPEA) can be expressed as:

$$X_N^{Bulk} = \min_{i=1,2,...(N-1)} \left\{ \alpha_{N(i)} \left( X_i^{Bulk} \right)^{-\beta_{N(i)}} e^{\frac{\Delta g_{N(i)}^{ppt.}}{kT}} \right\}. \qquad (58)$$

The maximum effects to reduce GB energy with increasing temperature for Type II HEGBs can be achieved for a specific bulk HEA in an equilibrium with a $(N-1)$ precipitates. Assuming, for simplicity, precipitates are $(N-1)$ binary intermetallic line compounds, the (optimal) composition of the HEA with maximum reduction of GB energy can be determined by the chemical potentials of the $(N-1)$ principates for $i = 1, 2, \ldots, (N-1)$ as:

$$X_N^{Bulk} = \alpha_{N(i)} \left( X_i^{Bulk} \right)^{-\beta_{N(i)}} e^{\frac{\Delta g_{N(i)}^{ppt.}}{kT}}. \tag{59}$$

This specific (optimized) HEA composition (for the maximum HEGB effects from GB adsorption of the $N$-th component) can be solved based on the above equation. The existence of ternary and high-order compounds or solution phases can also be considered for more realistic calculations.

To illustrated bulk high-entropy effects to reduce GB energy with increasing temperature and increasing number of components in Type II HEGBs, a simplified case of a fully saturated $(N-1)$-component ideal HEA solution with a minor strongly segregating $N$-th component ($X_N^{Bulk} \ll 1$) can be considered for a theoretical analysis. Here, the solvus line for each $i$-$N$ binary system can be expressed as $X_{N(i)}^{Binary\ Solvus} \approx \exp[\Delta g_{N(i)}^{ppt.}/(kT)]$, where $\Delta g_{N(i)}^{ppt.}$ is a constant (independent of $i$). As derived in Appendix III (with $\alpha_{N(i)} = 1$ for an ideal solution), the corresponding bulk solid solubility of the $N$-th component in the HEA limited by the precipitation of a binary line compound in the $i$-$N$ system is given by:

$$X_N^{Bulk} = \left( X_i^{Bulk} \right)^{-\beta_{N(i)}} e^{\frac{\Delta g_{N(i)}^{ppt.}}{kT}} = \left( X_i^{Bulk} \right)^{-\beta_{N(i)}} X_{N(i)}^{Binary\ Solvus}. \tag{60}$$

For a general case, the bulk solid solubility is given by $\min_{i=1,2,\ldots,(N-1)}\left[ \left( X_i^{Bulk} \right)^{-\beta_{N(i)}} X_{N(i)}^{Binary\ Solvus} \right]$ in a multicomponent HEA with $(N-1)$ principal components, if the $i$-$N$ binary systems have different thermodynamic properties. In other words, the bulk fraction of the segregating specie, $X_N^{Bulk}$, is determined (pinned) by the first $N$-enriched precipitate.

For a theoretical analysis to illustrate the key characteristics of Type II HEGBs, all $(N-1)$ binary subsystems are further assumed, for simplicity, to have identical thermodynamic properties. Thus: $\beta_{N(i)} = x/y \equiv \beta$, $\Delta g_{ads.(i)} = \Delta g_{ads.(P)}$, and $\Delta g_{N(i)}^{ppt.} = \Delta g_{N(P)}^{ppt.}$ (where "P" in the subscript represents Principal Component $i = 1, \ldots, N-1$) are taken to identical values for all $i \neq N$. In this simplified case, $(N-1)$ secondary phases will precipitate simultaneously. In this symmetric case with $X_N^{Bulk} \ll 1$, $X_i^{Bulk} = (1-X_N^{Bulk})/(N-1) \approx 1/(N-1)$ and for all $i \neq N$. Thus:

$$X_N^{Bulk} \approx \left( \frac{1}{N-1} \right)^{-\beta} e^{\frac{\Delta g_{N(i)}^{ppt.}}{kT}} = (N-1)^{\beta} e^{\frac{\Delta g_{N(i)}^{ppt.}}{kT}}. \tag{61}$$

Then, Equation (56) can be rewritten using the relation $\Delta g_{ads.(N \to P)} \equiv \Delta g_{ads.(N)} - \Delta g_{ads.(P)}$, as:

$$\begin{aligned}
\gamma_{GB} &= \overline{\gamma_{GB}^{(0)}} - \Gamma_0 kT \ln\left[\left(\sum_{i=1}^{N-1} X_i^{Bulk} e^{-\frac{\Delta g_{ads.(i)}}{kT}}\right) + X_N^{Bulk} e^{-\frac{\Delta g_{ads.(N)}}{kT}}\right] \\
&= \overline{\gamma_{GB}^{(0)}} - \Gamma_0 kT \ln\left[\left(1 - X_N^{Bulk}\right) e^{-\frac{\Delta g_{ads.(P)}}{kT}} + X_N^{Bulk} e^{-\frac{\Delta g_{ads.(N)}}{kT}}\right] \\
&= \overline{\gamma_{GB}^{(0)}} - \Gamma_0 kT \ln\left[e^{-\frac{\Delta g_{ads.(P)}}{kT}}\left(1 - X_N^{Bulk} + X_N^{Bulk} e^{-\frac{\Delta g_{ads.(N)}}{kT}} e^{\frac{\Delta g_{ads.(P)}}{kT}}\right)\right]. \quad (62)\\
&= \overline{\gamma_{GB}^{(0)}} + \Gamma_0 \Delta g_{ads.(P)} - \Gamma_0 kT \ln\left(1 - X_N^{Bulk} + X_N^{Bulk} e^{-\frac{\Delta g_{ads.(N \to P)}}{kT}}\right) \\
&\approx \overline{\gamma_{GB}^{(0)}} - \Gamma_0 kT \ln\left(1 + X_N^{Bulk} e^{-\frac{\Delta g_{ads.(N \to P)}}{kT}}\right) \quad (\text{if } X_N^{Bulk} \ll 1)
\end{aligned}$$

Here, $\Delta g_{ads.(P)}$ vanishes as $X_N^{Bulk} \to 0$ and $\exp[-\Delta g_{ads.(N \to P)}/(kT)] \gg 1$ for a strong segregating component $N$. In this case, $X_N^{Bulk}$ is maximized in the HEA saturated with the $N$-th component, with the composition moving along the multicomponent solvus line (in equilibrium with $(N - 1)$ principitates, which will appear simultaneously in this simplified symmetric system; noting that in a real asymmetric system, however, the first precipitate pins the chemical potential of the $N$-th component). Plugging Equation (61) into Equation (62) and using $\Delta g_{N(P)}^{ads.-ppt.} = \Delta g_{ads.(N \to P)} - \Delta g_{N(P)}^{ppt.}$ produce the following expression:

$$\gamma_{GB} \approx \overline{\gamma_{GB}^{(0)}} - \Gamma_0 kT \ln\left[1 + (N-1)^\beta e^{-\frac{\Delta g_{N(P)}^{ads.-ppt.}}{kT}}\right]. \quad (63)$$

Thus, the adsorption contribution to the effective GB entropy for Type II HEGBs is given by:

$$s_{GB}^{eff.\,(ads.,\,\text{saturated})} = -\left(\frac{d\gamma_{GB}}{dT}\right) \approx \Gamma_0 \left\{k \ln\left[f_{II}(\Delta g_{N(P)}^{ads.-ppt.}, N-1)\right] + \frac{\Delta g_{N(P)}^{ads.-ppt.}}{T}\left[1 - f_{II}(\Delta g_{N(P)}^{ads.-ppt.}, N-1)^{-1}\right]\right\} > 0, \quad (64)$$

where $f_{II}(\Delta g_{N(P)}^{ads.-ppt.}, N-1) \equiv 1 + (N-1)^\beta \exp[-\Delta g_{N(P)}^{ads.-ppt.}/(kT)]$ (so that $\gamma_{GB} \approx \overline{\gamma_{GB}^{(0)}} - \Gamma_0 kT \ln f_{II}$) is a dimensionless factor that increases with increasing $(N-1)$. Thus, $s_{GB}^{eff.\,(ads.,\,\text{saturated})}$ is positive and increases with increasing $(N-1)$, the number of principal elements in the HEA/MPEA.

Fig. 6 further shows an analysis of Type II HEGBs based on Equation (63) for a hypothetic saturated MPEA or HEA with $(N - 1)$ principal components plus one strongly segregating minor component. Here, it is assumed that the primary phase is in equilibrium with $(N - 1)$ simultaneously precipitated secondary phases of binary intermetallic compounds of identical $M_xS_y$ ($\beta = x/y$) stoichiometry and all binary subsystems have identical thermodynamic properties in this symmetric system for simplicity.

Computed GB energy reduction *vs.* temperature curves for $\Delta g_{N(P)}^{ads.-ppt.} = 0.1$ eV/atom and $\beta = 1, 2/3,$

and 3/2, respectively, are shown in Fig. 6. In all cases, GB energy decreases with increasing temperature (with a positive effective GB entropy) in the saturated HEAs and MPEAs. Furthermore, the effective GB entropy ($-d\gamma_{GB}/dT$) increases with increasing number of components in the MPEA/HEA primary phase. Here, the reduction in GB energy is also plotted in normalized parameter $(\gamma_{GB} - \overline{\gamma_{GB}^{(0)}})/\Gamma_0$ for generality. If the $\Gamma_0$ value for the NbMoTaW (110) twist GB in this lattice model is adopted, $(\gamma_{GB} - \overline{\gamma_{GB}^{(0)}})/\Gamma_0 = -0.18$ eV corresponds to an GB energy reduction of ~1.55 J/m² as a reference value, which represent a substantial reduction in GB energy.

An isothermal section of ternary phase diagram displaying the saturated composition for binary ($N - 1 = 1$) and ternary ($N - 1 = 2$) alloys are also shown in Fig. 6. This ternary phase diagram illustrates the increased solid solubility of the segregating element (always set to be the *N*-th component) with increasing number of the principal components in the primary bulk phase, which results in the increased GB adsorption of the *N*-th component. This mechanism is also expected to provide HEGB effects for $N > 3$. Here, the bulk high-entropy effect lowers the bulk chemical potentials (with respect to binary intermetallic precipitates) so that more adsorptions can be accommodated at GBs before the occurrence of precipitation.

In other words, HEGBs form from a competition between the GB adsorption and precipitation. If the precipitation is inhibited by the bulk high-entropy effects that lowers the chemical potentials of the HEA phase with respect to precipitation, it promotes the accommodation of GB adsorption to reduce GB energy.

It should be noted that this symmetric ideal solution model (with identical $\Delta g_{N(i)}^{ads.-ppt.}$ and $\beta_{N(i)}$ values for all binary subsystems) shows a special case with is no thermodynamic compositional degree of freedom according to the Gibbs phase rule, as the $(N-1)$ precipitated secondary phases pin the bulk chemical potential of the *N*-th component simultaneously. This is, however, not a general case. Assuming ideal mixing and $X_N^{Bulk} \ll 1$ for simplicity, for a system with different $\Delta g_{N(i)}^{ads.-ppt.}$ and/or different $\beta_{N(i)}$ values for different binary subsystems, Equation (63) can be generalized to:

$$\gamma_{GB} \approx \overline{\gamma_{GB}^{(0)}} - \Gamma_0 kT \ln \left\{ 1 + \min_{i=1,2,...(N-1)} \left[ \left( X_i^{Bulk} \right)^{-\beta_{N(i)}} e^{-\frac{\Delta g_{N(i)}^{ads.-ppt.}}{kT}} \right] \right\}. \tag{65}$$

Here, the chemical potential of the *N*-th component is pinned by the first precipitate of the *i-N* compound with the minimum value of $(X_i^{Bulk})^{-\beta_{N(i)}} \exp[-\Delta g_{N(i)}^{ads.-ppt.}/(kT)]$, which determines the actual $X_N^{Bulk}$ in the *N*-saturated HEA/MPEA phase.

Similar to Type I HEGBs, more realistic modeling of real HEAs with Type II HEGBs can be conducted by applying Equation (22) directly using the actual bulk solvus composition (the temperature-dependent solid solubility of the segregating *N*-th component) obtained by bulk CALPHAD methods. If the multicomponent solvus composition cannot be determined because of lacking thermodynamic data, the following approximation can be adopted to estimate the bulk solid solubility of the *N*-th component from binary solvus line (albeit the existence of the ternary and multicomponent precipitates, if any, can further limit the bulk solid solubility of the segregating *N*-th component):

$$X_N^{Bulk} \approx \min_{i=1,2,...(N-1)} \left[ \left( X_i^{Bulk} \right)^{-\beta_{N(i)}} X_{N(i)}^{Binary\ Solvus} \right]. \tag{66}$$

Again, in this case, $X_N^{Bulk}$ is different from (smaller than) the overall (initial) composition due to the precipitation. The actual bulk $X_i^{Bulk}$ ($i \neq N$) in the bulk phase should be determined based on the mass conservation considering the main and precipitation phases. If the fraction of secondary phase is negligibly small, it can be assumed that the relative ratios of principal elements remain unchanged after the precipitation. Otherwise, iterations can be used to quantify how precipitation of the secondary phase affects the bulk composition. Based on the above derived equations, one can use an ideal solution with the bonding energies and segregation enthalpies mimicking a real multicomponent alloy, along with real binary phase diagrams, to predict useful trends. Future studies should be conducted to examine these derived approximated expressions.

## C. General Cases and Discussion

A *N*-component alloy with *M* principal elements (Component 1, 2, … *M*) and *S* segregating solute elements (Component $M+1$, $M+2$, … $M+S$) can also be considered ($N = M + S$). To show the general characteristics and trends, let me start with a simple case of an ideal solution assuming: (1) $\sum_{j=M+1}^{M+S} X_j^{Bulk} \ll 1$, (2) *M* principal elements are equimolar ($X_i^{Bulk} \approx 1/M$ for $i \leq M$) with identical thermodynamic properties (*e.g.*, identical $\Delta g_{ads.(i)} = \Delta g_{ads.(P)}$ for simplicity), and (3) *S* minor segregating elements also have identical thermodynamic properties (*e.g.*, identical $\Delta g_{ads.(j)} = \Delta g_{ads.(S)}$ and identical β value and precipitation energy for all relevant binary systems; ignoring ternary and high-order compounds). Using Equation (22) or (41) and the relation $\Delta g_{ads.(P \to S)} \equiv \Delta g_{ads.(P)} - \Delta g_{ads.(S)}$, the following equation can be obtained:

$$\begin{aligned}
\gamma_{GB} &= \overline{\gamma_{GB}^{(0)}} - \Gamma_0 kT \ln\left( \sum_{i=1}^{M} X_i^{Bulk} e^{-\frac{\Delta g_{ads.(P)}}{kT}} + \sum_{j=M+1}^{M+S} X_j^{Bulk} e^{-\frac{\Delta g_{ads.(S)}}{kT}} \right) \\
&= \overline{\gamma_{GB}^{(0)}} - \Gamma_0 kT \ln\left[ e^{-\frac{\Delta g_{ads.(P)}}{kT}} \left( \sum_{i=1}^{M} X_i^{Bulk} + \sum_{j=M+1}^{M+S} X_j^{Bulk} e^{-\frac{\Delta g_{ads.(S)}}{kT}} e^{\frac{\Delta g_{ads.(P)}}{kT}} \right) \right] \\
&= \overline{\gamma_{GB}^{(0)}} + \Gamma_0 \Delta g_{ads.(P)} - \Gamma_0 kT \ln\left( \sum_{i=1}^{M} X_i^{Bulk} + \sum_{j=M+1}^{M+S} X_j^{Bulk} e^{-\frac{\Delta g_{ads.(S \to P)}}{kT}} \right) \\
&\approx \overline{\gamma_{GB}^{(0)}} - \Gamma_0 kT \ln\left( 1 + S \cdot X_i^{Bulk} \cdot e^{-\frac{\Delta g_{ads.(S \to P)}}{kT}} \right) \\
&= \overline{\gamma_{GB}^{(0)}} - \Gamma_0 kT \ln\left( 1 + S \cdot M^\beta \cdot e^{-\frac{\Delta g_{S(P)}^{ads.-ppt.}}{kT}} \right)
\end{aligned} \tag{67}$$

The above equation suggests strong HEGB effects to reduce GB energy with increasing temperature, where the GB energy reduction increases with increasing $S \cdot M^{\beta}$, where $M$ is the number of principal (solvent) elements and $S$ is the number of (minor) segregating solute elements. This is a more pronouncing effect thn Type I and Type II HEGBs.

Similar to the prior treatments, mixed HEGBs can also be modeled by applying Equation (22) directly using the actual bulk solvus composition obtained by bulk CALPHAD methods. If the multicomponent solvus compositions are not available, the following approximation can be adopted for the precipitated secondary phase that pins the chemical potential of the $j$-th component ($M+1 \leq j \leq M+S$) from binary solvus curve read in relevant binary phase diagrams (that are usually available):

$$X_j^{Bulk} \approx \min_{i=1,2,..M} \left[ \left( X_i^{Bulk} \right)^{-\beta_{j(i)}} X_{j(i)}^{Binary\ Solvus} \right]. \tag{68}$$

Similar to prior cases, $X_j^{Bulk}$ ($j > M$) is different from (smaller than) the overall fraction of the $j$-th component in the system due to the precipitation. The actual $X_i^{Bulk}$ ($i \leq M$) in the bulk phase should be determined based on the mass conservation. If the total fraction of secondary phases is negligibly small, it can be assumed that the relative ratios of principal elements remain unchanged after the precipitation.

As a final note, the total effective GB entropy can be greater than the GB adsorption contribution derived here for all types of HEGBs, with the additional contributions from GB disorder and GB free volume, as well as the coupling effects that possibly enhance both GB adsorption and GB disordering in a positive feedback loop.

## VI. CONCLUSIONS

In summary, GB segregation models have been derived for multicomponent alloys, particularly for HEAs and MPEAs. Differing from classical models where one component is taken as solvent and others are considered solutes, the models derived in this study are referenced to the bulk composition to enable improved treatments of HEAs and MPEAs with no principal component. An ideal solution model is first formulated and solved to obtain analytical expressions to predict GB segregation and GB energy. A regular solution model is further derived. The simplicity of the derived analytical expressions makes them useful for predicting trends. It is demonstrated that the GB composition calculated using the simple analytical expression derived in this study and data readily from the Materials Project agree well with a prior sphosipcated atomistic simulation, illustrating the predictability and usefulness of the simple analytical expressions.

As an application example of using the derived models to investigate nascent interfacial phenomena in compositionally complex GBs, the derived expressions are used to further formulate a set of equations to elucidate an emergent concept of HEGBs. Two types of HEGBs, as well as the mixed type, are analyzed, where HEGBs can be realized in saturated alloys with chemical potentials being pinned by precipitated secondary phases. Type I HEGBs can form in a saturated conventional multicomponent alloy, where ($N -$ 1) precipitated secondary phases pin the chemical potentials of the primary phase. Type II HEGBs form in HEAs or MPEAs with ($N - 1$) principal elements and one strongly segregating elements, where one or more precipitated secondary phases pin the bulk chemical potentials. Such effects can be further enhanced in the mixed Type I and II HEGBs, as shown by the derived model. These HEGBs possess three distinct characters: (1) GB energy reduces with increasing temperature, (2) the total GB adsorption amount increases with increasing temperature, and (3) the positive effective GB entropy increases with increasing number of components. The effects of interfacial structural disordering and GB free volume, which are not considered in the lattice models, can further increase the effective GB entropy. Coupling of interfacial disordering and GB adsorption may further enhance HEGB effects, which should be investigated in future studies.

## ACKNOWLEDGEMENTS

This work is supported by the U.S. Army Research Office (ARO Grant No. W911NF2210071, managed by Dr. Michael P. Bakas and Dr. Andrew D. Brown, in the Synthesis & Processing program). The author also acknowledges his former student Dr. Naixie Zhou (who started the HEGB analysis in his Ph.D. dissertation research in 2016 with the author) and a new Ph.D. student Keyu Cao for assistance in collecting some data, checking calculations, and proofreading.

## AUTHOR DECLARATIONS

**Conflict of Interest** The author has no conflicts to disclose.

## AUTHOR CONTRIBUTIONS

J.L. conceived the idea, conducted the derivations, calculations, and analysis, and wrote the manuscript.

## DATA AVAILABILITY

The data that support the findings of this study are available within the article.

## SUPPLEMENTARY MATERIAL

Three appendices of additional derivations are provided. No other supplementary material is needed or supplied.

**Appendix I:**

For a multicomponent alloy with one principal element (assumed to be the 1st component), let me define $r_1 \equiv X_1^{GB}/X_1^{Bulk}$ and re-write Equation (11) (assuming $j = 1$) as:

$$X_i^{GB} = r_1 X_i^{Bulk} e^{-\frac{\Delta h_{ads.(i \to 1)}}{kT}}. \tag{69}$$

Subsequently, $r_1$ can be determined by $\sum_{i>1} X_i^{GB} = 1 - X_1^{GB} = 1 - r_1 X_1^{Bulk}$, so that:

$$r_1 = \frac{1}{X_1^{Bulk} + \sum_{j>1} X_j^{Bulk} e^{-\frac{\Delta h_{ads.(j \to 1)}}{kT}}}. \tag{70}$$

and

$$X_i^{GB} = \frac{X_i^{Bulk} e^{-\frac{\Delta h_{ads.(i \to 1)}}{kT}}}{X_1^{Bulk} + \sum_{j>1} X_j^{Bulk} e^{-\frac{\Delta h_{ads.(j \to 1)}}{kT}}}. \tag{71}$$

An expression similar to Equation (17) can be obtained as:

$$\Delta h_{ads.(i \to 1)} + kT \ln\left(\frac{X_i^{GB}}{X_i^{Bulk}}\right) = kT \ln r_1, \tag{72}$$

which holds for any component $i$ ($\neq 1$). Assuming $j = 1$ in Equation (10) and plugging $X_1^{GB} = 1 - \sum_{i>1} X_i^{GB}$ and the above equations, the following expression can be obtained:

$$\begin{aligned}
\gamma_{GB} &= \gamma_{GB,1}^{(0)} - \Gamma_0 X_1^{GB} \Delta E_1^{strain} + \Gamma_0 X_1^{GB} kT \ln\left(\frac{X_i^{GB}}{X_i^{Bulk}}\right) + \Gamma_0 \sum_{i>1} X_i^{GB} \left[\tfrac{1}{2} Q z_v \left(|e_{ii}| - |e_{11}|\right) - \Delta E_i^{strain}\right] \\
&= \gamma_{GB,1}^{(0)} - \Gamma_0 \left(1 - \sum_{i>1} X_i^{GB}\right) \Delta E_1^{strain} + \Gamma_0 X_1^{GB} kT \ln\left(\frac{X_i^{GB}}{X_i^{Bulk}}\right) + \Gamma_0 \sum_{i>1} X_i^{GB} \left[\tfrac{1}{2} Q z_v \left(|e_{ii}| - |e_{11}|\right) - \Delta E_i^{strain}\right] \\
&= \gamma_{GB,1}^{(0)} - \Gamma_0 \Delta E_1^{strain} + \Gamma_0 X_1^{GB} kT \ln\left(\frac{X_i^{GB}}{X_i^{Bulk}}\right) + \Gamma_0 \sum_{i>1} X_i^{GB} \left[\tfrac{1}{2} Q z_v \left(|e_{ii}| - |e_{11}|\right) - \left(\Delta E_i^{strain} - \Delta E_1^{strain}\right)\right] \\
&= \gamma_{GB,1}^{(0)} - \Gamma_0 \Delta E_1^{strain} + \Gamma_0 X_1^{GB} kT \ln\left(\frac{X_i^{GB}}{X_i^{Bulk}}\right) + \Gamma_0 \sum_{i>1} X_i^{GB} \left(\Delta h_{ads.(i \to 1)} + kT \ln\left(\frac{X_i^{GB}}{X_i^{Bulk}}\right)\right) \\
&= \gamma_{GB,1}^{(0)} - \Gamma_0 \Delta E_1^{strain} + \Gamma_0 X_1^{GB} kT \ln r_1 + \Gamma_0 \sum_{i>1} X_i^{GB} kT \ln r_1 \\
&= \gamma_{GB,1}^{(0)} - \Gamma_0 \Delta E_1^{strain} + \Gamma_0 \sum_i X_i^{GB} kT \ln r_1 \\
&= \left(\gamma_{GB,1}^{(0)} - \Gamma_0 \Delta E_1^{strain}\right) - \Gamma_0 kT \ln\left(X_1^{Bulk} + \sum_{i>1} X_i^{Bulk} e^{-\frac{\Delta h_{ads.(i \to 1)}}{kT}}\right)
\end{aligned} \tag{73}$$

Here, $\Delta E_1^{strain} \to 0$ as $X_1^{Bulk} \to 1$. While an analytical solution is also obtained here using the 1st component as a reference, Equation (22) using the bulk composition as a reference is a better option for HEAs.

**Appendix II:**

The multicomponent regular-solution model derived in this study is an extension of the Wynblatt *et al.*'s model[6,7] for a binary M-S alloy (M = matrix element and S = solute element; $X_S^{Bulk} + X_M^{Bulk} = 1$ and $X_S^{GB} + X_M^{GB} = 1$), in which the pair-interaction contribution to the adsorption enthalpy for $i \leq J_{Max}$ is given by:

$$\Delta h_{ads.(S \to M)}^{(\omega)} = 2\omega \left[ zX_S^{Bulk} - z^i X_S^i - \sum_{j=1}^{J_{Max}} z^j X_S^{i+j} - \sum_{j=1}^{i-1} z^j X_S^{i-j} - (1-Q) \sum_{j=i}^{J_{Max}} z^j X_S^i - \tfrac{1}{2} Q \sum_{j=i}^{J_{Max}} z^j \right], \quad (74)$$

Considering the bilayer adsorption in a twist GB and setting $J_{max} = 1$, $i = 1$, $X_B^1 = X_B^{GB}$, and $X_B^2 = X_B^{Bulk}$, the above equation can be rewritten as:

$$\begin{aligned}
\Delta h_{ads.(S \to M)}^{(\omega)} &= 2\omega \left[ zX_S^{Bulk} - z_l X_S^{GB} - z_v X_S^{Bulk} - (1-Q) z_v X_S^{GB} - \tfrac{1}{2} Q z_v \right] \\
&= 2\omega \left[ (z - z_v)(X_S^{Bulk} - X_S^{GB}) + Q z_v (X_S^{GB} - \tfrac{1}{2}) \right] \\
&= \omega \left[ 2(z - z_v)(X_S^{Bulk} - X_S^{GB}) \right] + Q z_v \omega (X_S^{GB} - X_M^{GB}) \\
&= \left[ \omega(z - z_v)(X_S^{Bulk} - X_S^{GB}) - \omega Q z_v X_M^{GB} \right] - \left[ \omega(z - z_v)(X_M^{Bulk} - X_M^{GB}) - \omega Q z_v X_S^{GB} \right] \\
&= \Delta h_{ads.(S)}^{(\omega)} - \Delta h_{ads.(M)}^{(\omega)}
\end{aligned} \quad (75)$$

The above derivation shows that the case of $N = 2$ of the multicomponent regular solution model derived in this study is equivalent to a case of bilayer adsorption at a twist GB (with $J_{max} = 1$) in Wynblatt *et al.*'s model[6,7] for binary alloys. This equivalence also verifies the derivation of the multicomponent regular solution model presented in this work, which can be considered as a generalization of Wynblatt *et al.*'s binary regular solution model[6,7] to multicomponent regular solutions.

**Appendix III:**

Considering a binary $M_xS_y$ line compound (M = matrix element and S = solute element), its formation Gibbs free energy is defined by:

$$x\mu_M + y\mu_S = x\mu_M^{(0)} + y\mu_S^{(0)} + \Delta g_{formation}. \tag{76}$$

For a saturated multicomponent solution in an equilibrium with this $M_xS_y$ line compound (where the chemical potential of the solute is pinned by this compound), the Gibbs formation free energy can be expressed as:

$$\begin{aligned}\Delta g_{formation} &= x\left(\mu_M - \mu_M^{(0)}\right) + y\left(\mu_S - \mu_S^{(0)}\right) \\ &= kT\left[x\ln(\gamma_M X_M) + y\ln(\gamma_S X_S)\right]\end{aligned}. \tag{77}$$

The above equation can be re-written as:

$$(X_S)^y (\gamma_S)^y (\gamma_M)^x (X_M)^x = e^{\frac{\Delta g_{formation}}{kT}}. \tag{78}$$

Thus:

$$X_S = (\gamma_S)^{-1}(\gamma_M)^{-x/y}(X_M)^{-x/y} e^{\frac{\Delta g_{formation}/y}{kT}} = \alpha_{S(M)}(X_M)^{-\beta_{S(M)}} e^{\frac{\Delta g_{S(M)}^{ppt.}}{kT}}, \tag{79}$$

where:

$$\begin{cases}\Delta g_{S(M)}^{ppt.} \equiv \Delta g_{formation}/y \\ \beta_{S(M)} \equiv x/y \\ \alpha_{S(M)} \equiv (\gamma_M)^{-\beta_{S(M)}}(\gamma_S)^{-1}\end{cases}. \tag{80}$$

Here, $\Delta g_{S(M)}^{ppt.} \approx \Delta h_{S(M)}^{ppt.}$ (< 0) is negative here. For an ideal solution, $\alpha_{S(M)} = 1$ so that $X_S = (X_M)^{-\beta_{S(M)}}\exp[\Delta g_{S(M)}^{ppt.}/(kT)]$. For a binary dilute solution obeying Henry's law, $\alpha_{S(M)} = (\gamma_S)^{-1}$ is a constant ($X_0$) and $(X_M)^{\beta_{S(M)}} \approx 1$ so that $X_S \approx X_0 \exp[\Delta g_{S(M)}^{ppt.}/(kT)]$. For a dilute ideal solution, $X_S \approx \exp[\Delta g_{S(M)}^{ppt.}/(kT)]$.

Equation (77) and its simplification for ideal solutions, $X_S = (X_M)^{-\beta_{S(M)}}\exp[\Delta g_{S(M)}^{ppt.}/(kT)]$, can be extended to the case where a multicomponent solution is in an equilibrium with a binary line compound, which are also be used the derivation in the main text.

**References:**

1. D. McLean, Oxford, Clarendon Press (1957).

2. R. H. Fowler and E. A. Guggenheim, *Stastistical Thermodynmaics* (Cambridge University Press, 1939).

3. E. D. Hondros and M. P. Seah, Metallurgical Transactions **8A,** 1363 (1977).

4. P. Lejcek and S. Hofmann, Critical Reviews in Solid State and Materials Sciences **33,** 133 (2008).

5. P. Wynblatt and R. Ku, Surface Science **65,** 511 (1977).

6. P. Wynblatt and D. Chatain, Metallurgical and Materials Transactions a-Physical Metallurgy and Materials Science **37A,** 2595 (2006).

7. P. Wynblatt, D. Chatain, and Y. Pang, Journal of Materials Science **41,** 7760 (2006).

8. P. Lejcek, *Grain boundary segregation in metals*, Vol. 136 (Springer Science & Business Media, 2010).

9. M. Guttmann, Surface Science **53,** 213 (1975).

10. B. Cantor, I. Chang, P. Knight, and A. Vincent, Materials Science and Engineering: A **375,** 213 (2004).

11. J. W. Yeh, S. K. Chen, S. J. Lin, J. Y. Gan, T. S. Chin, T. T. Shun, C. H. Tsau, and S. Y. Chang, Advanced Engineering Materials **6,** 299 (2004).

12. E. P. George, D. Raabe, and R. O. Ritchie, Nature Reviews Materials **4,** 515 (2019).

13. D. B. Miracle and O. N. Senkov, Acta Materialia **122,** 448 (2017).

14. P. Wynblatt and D. Chatain, Physical Review Materials **3,** 054004 (2019).

15. C. Hu and J. Luo, Materials Horizons **9,** 1023 (2022).

16. L. Li, R. D. Kamachali, Z. Li, and Z. Zhang, Physical Review Materials **4,** 053603 (2020).

17. L. Wang and R. D. Kamachali, Journal of Alloys and Compounds **933,** 167717 (2023).

18. N. Zhou, T. Hu, J. Huang, and J. Luo, Scripta Materialia **124,** 160 (2016).

19. J. Luo and N. Zhou, Communications Materials **4,** 7 (2023).

20. X.-G. Li, C. Chen, H. Zheng, Y. Zuo, and S. P. Ong, npj Computational Materials **6,** 70 (2020).

21. P. Wynblatt and D. Chatain, Materials Science and Engineering a-Structural Materials Properties Microstructure and Processing **495,** 119 (2008).

22. J. Friedel, Advances in Physics **3,** 446 (1954).

23. M. Guttmann and D. McLean, (ASM, Metals Park, 1979).

24. N. Zhou, C. Hu, and J. Luo, Acta Materialia **221,** 117375 (2021).

25. N. X. Zhou, T. Hu, and J. Luo, Current Opinion in Solid State & Materials Science **20,** 268 (2016).

26. P. Lejček, M. Všianská, and M. Šob, Journal of Materials Research **33,** 2647 (2018).

27 M. Černý, P. Šesták, M. Všianská, and P. Lejček, Metals **12,** 1389 (2022).

28 P. Lejček, S. Hofmann, M. Všianská, and M. Šob, Acta Materialia **206,** 116597 (2021).

29 A. Jain, S. P. Ong, G. Hautier, W. Chen, W. D. Richards, S. Dacek, S. Cholia, D. Gunter, D. Skinner, and G. Ceder, APL materials **1** (2013).

30 N. Zhou, T. Hu, and J. Luo, Current Opinion in Solid State and Materials Science **20,** 268 (2016).

31 J. M. Rickman and J. Luo, Current Opinion in Solid State and Materials Science **20,** 225 (2016).

32 P. R. Cantwell, M. Tang, S. J. Dillon, J. Luo, G. S. Rohrer, and M. P. Harmer, Acta Materialia **62,** 1 (2014).

33 J. Luo, Applied Physics Letters **95,** 071911 (2009).

34 J. Luo, Interdisciplinary Materials **2,** in press (2023).

35 J. Luo, Current Opinion in Solid State and Materials Science **12,** 81 (2008).

36 J. Luo, Critical Reviews in Solid State and Material Sciences **32,** 67 (2007).

37 C. Hu and J. Luo, Scripta Materialia **158,** 11 (2019).

38 S. Yang, N. Zhou, H. Zheng, S. P. Ong, and J. Luo, Physical review letters **120,** 085702 (2018).

39 P. R. Cantwell, T. Frolov, T. J. Rupert, A. R. Krause, C. J. Marvel, G. S. Rohrer, J. M. Rickman, and M. P. Harmer, Annual Review of Materials Research **50,** 465 (2020).

40 T. Meiners, T. Frolov, R. E. Rudd, G. Dehm, and C. H. Liebscher, Nature **579,** 375 (2020).

41 F. Cao, Y. Chen, S. Zhao, E. Ma, and L. Dai, Acta Materialia **209,** 116786 (2021).

42 T. Frolov, D. L. Olmsted, M. Asta, and Y. Mishin, Nature communications **4,** 1899 (2013).

43 T. Frolov, M. Asta, and Y. Mishin, Physical Review B **92,** 020103 (2015).

44 M. Tang, W. C. Carter, and R. M. Cannon, Physical Review B **73,** 024102 (2006).

45 S. J. Dillon, M. Tang, W. C. Carter, and M. P. Harmer, Acta Materialia **55,** 6208 (2007).

46 M. Tang, W. C. Carter, and R. M. Cannon, Physical review letters **97,** 075502 (2006).

47 Y. Mishin, W. Boettinger, J. Warren, and G. McFadden, Acta Materialia **57,** 3771 (2009).

48 J. Luo, V. Gupta, D. Yoon, and H. Meyer III, Applied Physics Letters **87,** 231902 (2005).

49 X. Shi and J. Luo, Applied Physics Letters **94,** 251908 (2009).

50 T. J. Rupert, Current Opinion in Solid State and Materials Science **20,** 257 (2016).

51 C. M. Grigorian and T. J. Rupert, Journal of Materials Research **37,** 554 (2022).

52 J. Weissmüller, Nanostructured Materials **3,** 261 (1993).

53 J. Weissmuller, Journal of Materials Research **9,** 4 (1994).

54 C. C. Koch, Journal of Materials Science **42,** 1403 (2007).

55. K. A. Darling, R. N. Chan, P. Z. Wong, J. E. Semones, R. O. Scattergood, and C. C. Koch, Scripta Materialia **59,** 530 (2008).
56. C. C. Koch, R. O. Scattergood, K. A. Darling, and J. E. Semones, Journal of Materials Science **43,** 7264 (2008).
57. N. Zhou and J. Luo, Materials Letters **115,** 268 (2014).
58. A. R. Kalidindi and C. A. Schuh, Acta Materialia **132,** 128 (2017).
59. A. R. Kalidindi, T. Chookajorn, and C. A. Schuh, JOM **67,** 2834 (2015).
60. H. A. Murdoch and C. A. Schuh, Acta Materialia **61,** 2121 (2013).
61. T. Chookajorn, H. A. Murdoch, and C. A. Schuh, Science **337,** 951 (2012).
62. J. R. Trelewicz and C. A. Schuh, Physical Review B **79** (2009).
63. J. W. Cahn, Acta Metallurgica et Materialia **10,** 789 (1962).
64. S. M. Foiles, Scripta Materialia **62,** 231 (2010).
65. M. P. Seah, Journal of Physics F: Metal Physics **10,** 1043 (1980).
66. M. Qin, S. Shivakumar, and J. Luo, Journal of Materials Science **58,** 8548 (2023).
67. X. C. Liu, H. W. Zhang, and K. Lu, Science **342,** 337 (2013).
68. A. J. Detor and C. A. Schuh, Journal of Materials Research **22,** 3233 (2007).

**Table 1.** Parameters and calculated results for the equimolar NbMoTaW. All input data were taken from the Materials Project.[29] Bond energies were calculated from the $E_0$ (= $-\frac{1}{2}z\varepsilon_{ii}$) fitted for the Mie-Gruneisen equation of state. Bulk and shear moduli using the Voigt-Reuss-Hill approximation were adopted. Note that the mean is identical to the weighted average value for an equimolar HEA. The calculations were performed for a large-angle (110) twist GB ($z_v$ = 2 and $Q$ = 1/6 to represent a general GB). Using all data from density functional calculation (DFT) calculations in the Materials Project offers a consistent and accessible method for parameterization of the model.

| | Nb | Mo | Ta | W | (Weighted) Mean |
|---|---|---|---|---|---|
| Bulk Composition $X_i^{Bulk}$ | 0.25 | 0.25 | 0.25 | 0.25 | |
| Lattice Parameter $a$ (nm) <br> Density of GB Sites $\Gamma_0 = 2/(\sqrt{2}\bar{a}^2)\times 2$ | 0.332 | 0.317 | 0.331 | 0.317 | $\bar{a}$ = 0.324 nm <br> $\Gamma_0$ = 26.9 atoms/nm² |
| Radius $r_i = a/(2\sqrt{3})$ (nm) | 0.144 | 0.137 | 0.143 | 0.137 | $\bar{r}$ = 0.140 nm |
| Bulk Modulus $K$ (GPa) | 172 | 263 | 195 | 302 | |
| Shear Modulus | 26 | 117 | 65 | 148 | $\bar{G}$ = 89 GPa |
| Strain Energy $\Delta E_i^{strain} = -\Delta h_{ads.(i)}^\varepsilon$ (eV/atom) | 0.0132 | 0.0132 | 0.0105 | 0.0138 | |
| Bond Energy $\varepsilon_{ii}$ (eV/bond) | -2.52 | -2.71 | -2.96 | -3.24 | $\overline{\varepsilon_{self}}$ = -2.86 eV |
| $\Delta h_{ads.(i)}^\gamma = \frac{1}{2}Qz_v\left(|e_{ii}|-|\overline{e_{self}}|\right)$ (eV/atom) | -0.0560 | -0.0246 | +0.0173 | +0.0634 | |
| $\Delta h_{ads.(i)}^{(0)} = \Delta h_{ads.(i)}^\gamma + \Delta h_{ads.(i)}^\varepsilon$ (eV/atom) | -0.0692 | -0.0378 | +0.0068 | +0.0496 | |
| $\gamma_{GB,i}^{(0)} = \Gamma_0\left(\frac{1}{2}Qz_v|e_{ii}|\right)$ (eV/nm²) | 11.31 | 12.16 | 13.29 | 14.53 | $\overline{\gamma_{GB}^{(0)}}$ = 12.82 eV/nm² |
| $\gamma_{GB,i}^{(0)}$ (J/m²) | 1.81 | 1.95 | 2.13 | 2.33 | $\overline{\gamma_{GB}^{(0)}}$ = 2.05 J/m² |

**Table 2.** Calculated results for the non-equimolar Nb$_{0.155}$Mo$_{0.246}$Ta$_{0.280}$W$_{0.319}$. The same input data from the Materials Project[29] were used and the calculations were also performed for a large-angle (110) twist GB ($z_v$ = 2 and $Q$ = 1/6 to represent a general large-angle GB). Individual lattice parameters, radii, bulk and shear moduli, and self-bonding energies for Nb, Mo, Ta, and W are listed in Table 1. The weighted average radius, modulus, self-bonding energy, as well as the strain energies and segregation enthalpies, which depend on the bulk composition, are shown here. The computed GB composition at 300 K for this non-equimolar Nb$_{0.155}$Mo$_{0.246}$Ta$_{0.280}$W$_{0.319}$ alloy is shown in Table 3 to compare with atomistic simulation results.

| | **Nb** | **Mo** | **Ta** | **W** |
|---|---|---|---|---|
| Bulk Composition $X_i^{Bulk}$ | 0.155 | 0.246 | 0.280 | 0.319 |
| Strain Energy $\Delta E_i^{strain} = -\Delta h_{ads.(i)}^{\varepsilon}$ (eV/atom) | 0.0178 | 0.0105 | 0.0147 | 0.0109 |
| $\Delta h_{ads.(i)}^{\gamma} = \frac{1}{2} Q z_v \left( |e_{ii}| - |\overline{e_{self}}| \right)$ (eV/atom) | -0.0663 | -0.0349 | +0.0069 | +0.0531 |
| $\Delta h_{ads.(i)}^{(0)} = \Delta h_{ads.(i)}^{\gamma} + \Delta h_{ads.(i)}^{\varepsilon}$ (eV/atom) | -0.0841 | -0.0454 | -0.0077 | +0.0421 |
| Weighted Mean Lattice Parameter | $\overline{a}$ = 0.323 nm ||||
| Density of GB Sites | $\Gamma_0$ = 27.1 atoms/nm$^2$ ||||
| Weighted Mean Radius | $\overline{r}$ = 0.140 nm ||||
| Weighted Mean Shear Modulus | $\overline{G}$ = 98 GPa ||||
| Weighted Mean Bond Energy | $\overline{\varepsilon_{self}}$ = -2.92 eV ||||
| Weighted Mean $\gamma_{GB,i}^{(0)}$ | $\overline{\gamma_{GB}^{(0)}}$ = 13.18 eV/nm$^2$ or 2.11 J/m$^2$ ||||

**Table 3.** Comparison of the results from the analytical solution derived in this work with hybrid Monte Carlo and molecular dynamics (MC/MD) simulation of a polycrystal at 300 K in a supercell with six randomly inserted GBs and average grain diameter of ~7.5 nm conducted by Li *et al.*[20]. Note that bulk (grain) composition deviated from the equimolar composition due to GB segregation and a small grain size in a closed system in the MC/MD simulation. To do a fair comparison, The values of $X_i^{GB}$ were calculated using the analytical expression based on the non-equimolar bulk composition (Nb$_{0.155}$Mo$_{0.246}$Ta$_{0.280}$W$_{0.319}$) matching that observed in the MC/MD simulation. The calculated results for the (110) twist GB ( $z_v$ = 2 and $Q$ =1/6; representing a general large-angle GB with moderate segregation) using a simple analytical expression derived in this study agree well with the MC/MD simulation.

|  |  | **Nb** | **Mo** | **Ta** | **W** |
|---|---|---|---|---|---|
| MC/MD Simulation by Li *et al.* [20] | (Observed) Bulk Composition $X_i^{Bulk}$ | 0.155 | 0.246 | 0.280 | 0.319 |
|  | Average $X_i^{GB}$ in a Polycrystal | 0.577 | 0.267 | 0.136 | 0.020 |
| Calculations Using an Analytical Expression Derived in This Study | (Set) Bulk Composition $X_i^{Bulk}$ | 0.155 | 0.246 | 0.280 | 0.319 |
|  | $X_i^{GB}$ for a Large-Angle (110) Twist GB | 0.683 | 0.242 | 0.064 | 0.011 |

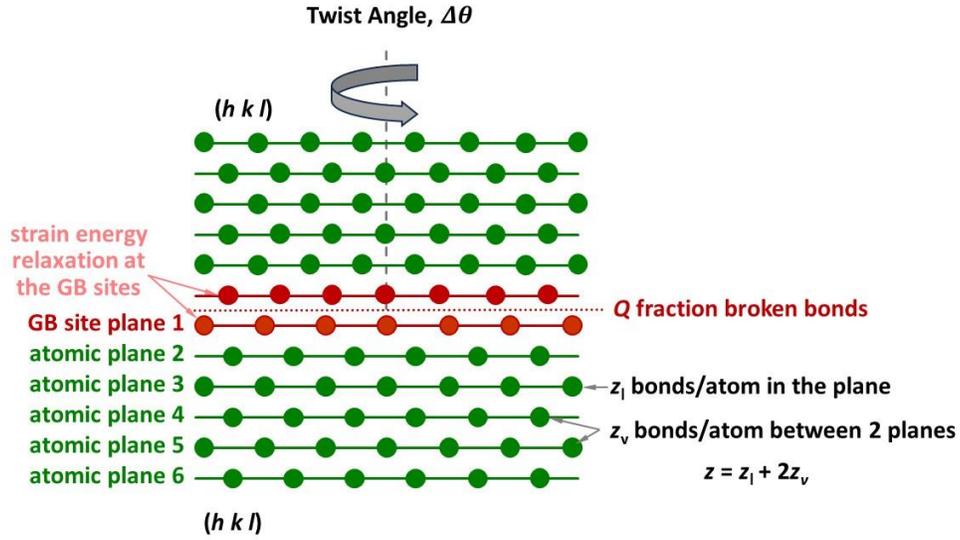

**FIG. 1.** Schematic illustration of the lattice model for a twist grain boundary (GB), which is adapted from the Wynblatt *et al.*'s model for a twist GB with $J_{max} = 1$ in a binary alloy,[21] but generalized for a multicomponent alloy (including HEAs and MPEAs) here. Each atom has $z$ bonds ($z = z_l + 2z_v$), including $z_l$ lateral bonds within the atomic layer and $z_v$ bonds between two neighbor atomic layers parallel to the twist GB. It is assumed that $Q$ fraction of the bonds is broken at the GB core between the two GB layers of the twist GB ($Q$ = 1/6 to represent a general large-angle GB to ensure $\gamma_{GB}/\gamma_{surface} = \frac{1}{3}$ to match typical experimental values). It is assumed, for simplicity, the adsorption is limited at the two GB layers (while acknowledging multilayer adsorption can take place in strong segregation region in regular solutions [21,24]). The differential bonding energies (due to broken bonds) and strain energy relaxation at the GB sites can drive GB adsorption (*a.k.a.* segregation).

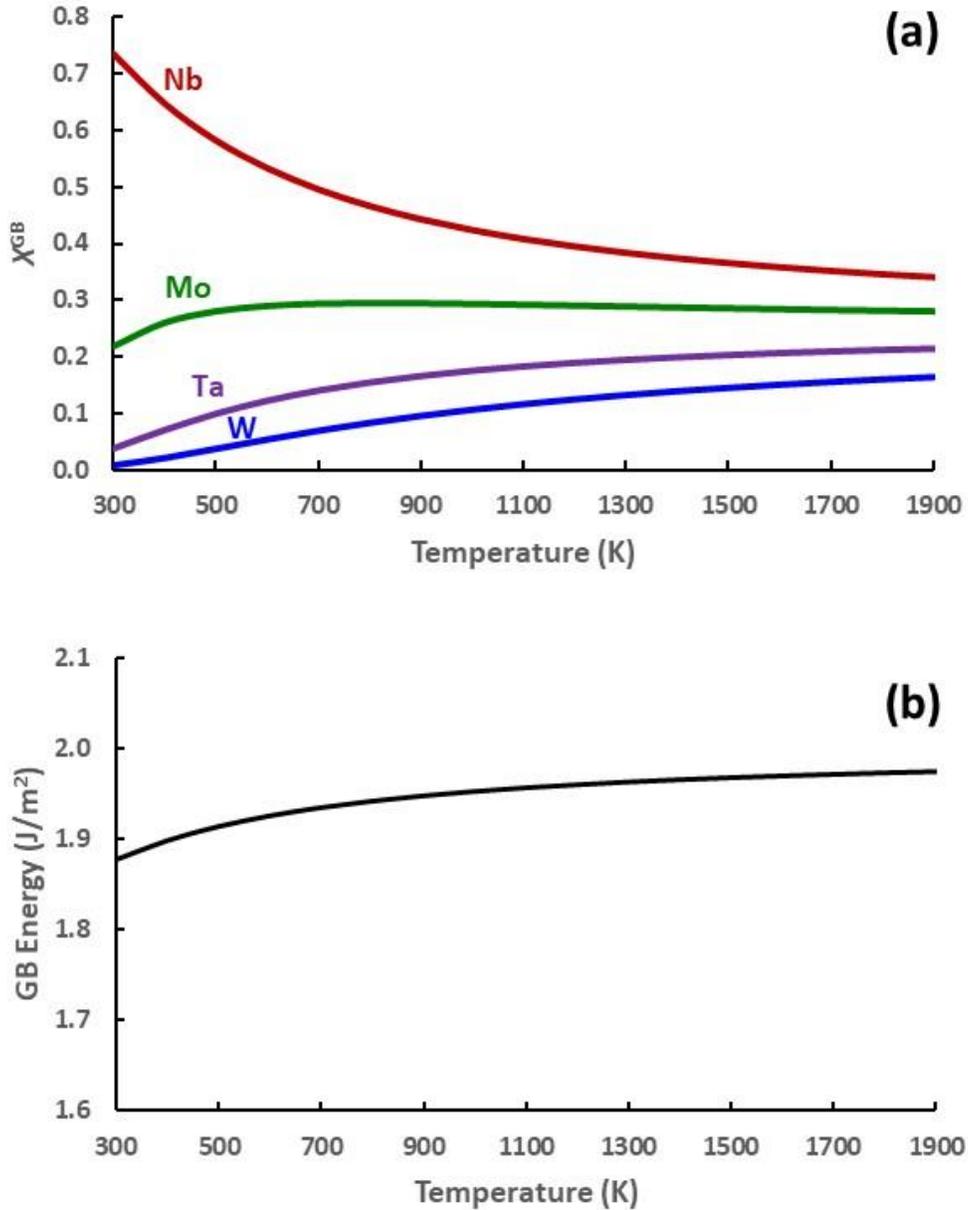

**FIG. 2.** Computed **(a)** GB composition $X_i^{GB}$ and **(b)** GB energy $\gamma_{GB}$ versus temperature curves for a high-angle (110) twist GB ($z_v$ = 2 and $Q$ = 1/6 to represent a general GB with moderate segregation tendency) for an equimolar NbMoTaW, using the analytical expressions derived in this study. All input data were from DFT calculations in the Materials Project[29] to offer consistency and convenience for parameterization (while noting that bond energies estimated from DFT calculations are typically higher than those calculated from atomization enthalpies in prior studies, so the calculated GB energies are higher). For a fixed bulk composition, GB energy decreases with increasing temperature due to temperature-induced desorption. The current model does not consider interfacial disordering and GB free volume, which can reduce the GB energy with increasing temperature.

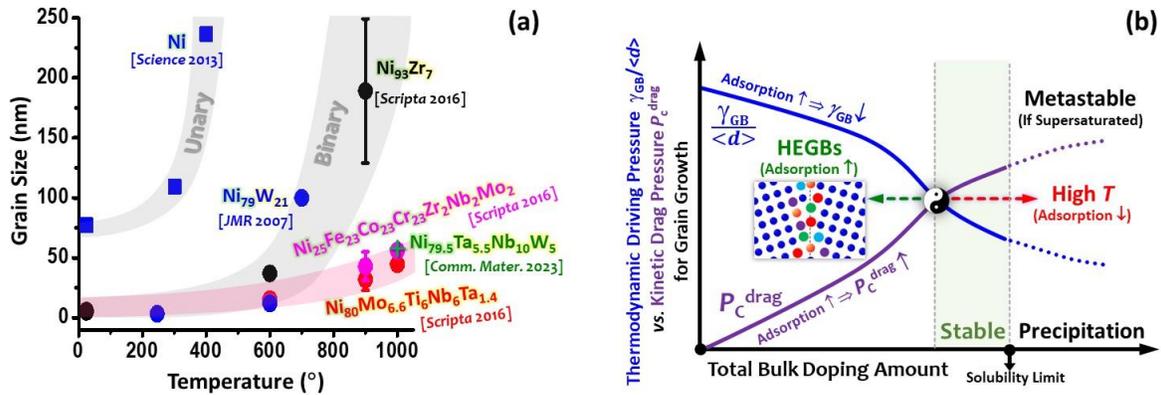

**FIG. 3. (a)** Prior experiments[18,19] showing the stabilization of three nanoalloys with high-entropy grain boundaries (HEGBs) against grain growth at high temperatures, in comparison with nanocrystalline unary and binary nanoalloys.[18,67,68] **(b)** A proposed mechanism[19] of stabilizing a nanoalloy against grain growth at high temperatures via HEGBs. To stabilize a nanoalloy, reduced thermodynamic driving pressure and increased critical kinetic solute drag pressure for grain growth has to achieve a balance below the solid solubility limit. Increasing temperature can de-stabilize the nanoalloy by inducing GB desorption, while HEGBs can counter this effect via increasing total adsorption with increasing number of components.[19] It was proposed[19] that HEGBs can simultaneously reduce the thermodynamic driving force and increase the kinetic solute drag, thereby increasing the high-temperature stability of nanoalloys. This figure is reprinted from a perspective article by Luo and Zhou, *Communications Materials* 2023, 4, 7 (© The Author(s) 2023 under a Creative Commons Attribution 4.0 International License).[19]

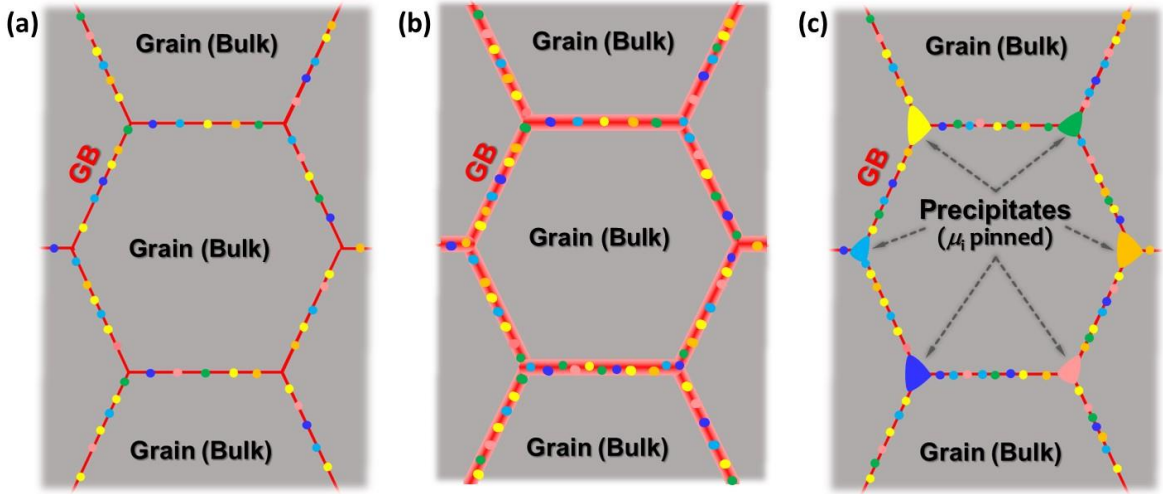

**FIG. 4.** Schematic illustration of GB segregation (*a.k.a.* interfacial adsorption in thermodynamics) in multicomponent alloys in three scenarios with different influences on effective GB entropies. **(a)** For a single-phase multicomponent alloy with a fixed bulk composition and a fixed number of GB adsorption sites, GB energy should decrease with increasing temperature due to temperature-induced desorption, so that the effective GB entropy is negative (without considering the effects of interfacial disordering and GB free volume, which can get rise to positive effective GB entropy). **(b)** Interfacial disordering can decrease GB energy with increasing temperature to produce a positive effective GB entropy in a unary system.[64] In multicomponent systems, interfacial disordering (and widening), particularly the formation of liquid-like GBs,[34,36] can enhance GB adsorption. The coupling of interfacial disordering and adsorption may subsequently promote the formation of HEGBs, but this hypothesis has not been rigorously approved yet. **(c)** Type I HEGBs can form from GB adsorption effects in saturated alloys with $(N-1)$ precipitated secondary phases that pin the chemical potentials of the primary phase, where GB energy decreases with increasing temperature and effective GB entropy increases with the increasing number of components. A further analysis shows the Type II HEGBs of similar characters can also form in HEAs saturated with one or more precipitated secondary phase(s) that pins the chemical potential of the segregation element(s) in the (primary) high-entropy phase. The coupling of Scenarios (c) and (b) may further increase the effective GB entropy, pointing to a future exploration direction.

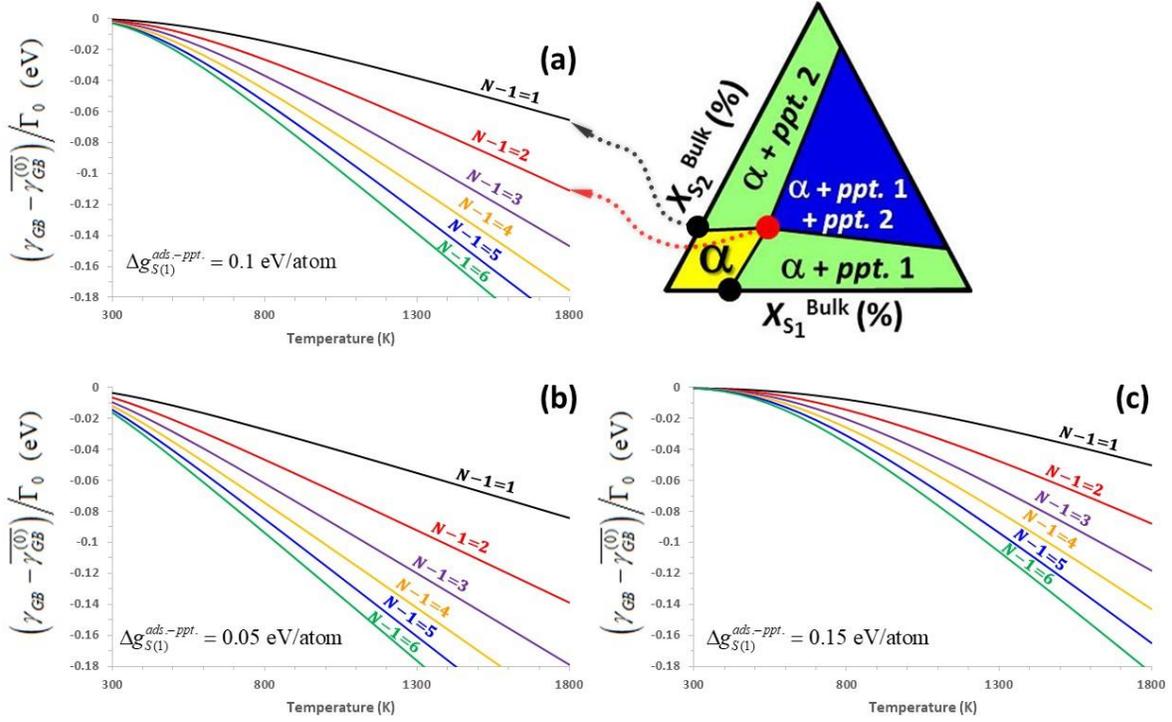

**FIG. 5.** Analysis of Type I HEGBs in normalized parameters based on the Equation (53) for a hypothetic symmetric saturated multicomponent alloy with one principal component and $(N-1)$ segregating minor components, with $(N-1)$ precipitated secondary phases that pin the chemical potentials. Computed GB energy reduction *vs.* temperature curves for $\Delta g_{S(1)}^{ads.-ppt.}$ = **(a)** 0.1 eV/atom, **(b)** 0.05 eV/atom, and **(a)** 0.15 eV/atom, respectively. In all cases, GB energy decreases with increasing temperature (producing a positive effective GB entropy) in the saturated multicomponent alloy, where the chemical potentials are pined by the $(N-1)$ precipitated secondary phases. Furthermore, the effective GB entropy ($-d\gamma_{GB}/dT$) increases with increasing number of components. If $\Gamma_0$ value for the Ni (100) twist GB is adopted as a reference, $(\gamma_{GB}-\overline{\gamma_{GB}^{(0)}})/\Gamma_0 = -0.18$ eV corresponds to an GB energy reduction of ~0.969 J/m². An isothermal section of ternary phase diagram displaying the saturated composition for binary ($N-1=1$) and ternary ($N-1=2$) alloys are also shown. This ternary phase diagram illustrates the increased total solid solubility of all solutes in the primary bulk phase from a binary alloy (represented by the two black dots) to a ternary alloy (represented by the red dot), which results in the increased total GB adsorption to reduce GB energy. The same mechanism is also expected to provide HEGB effects for $N > 3$.

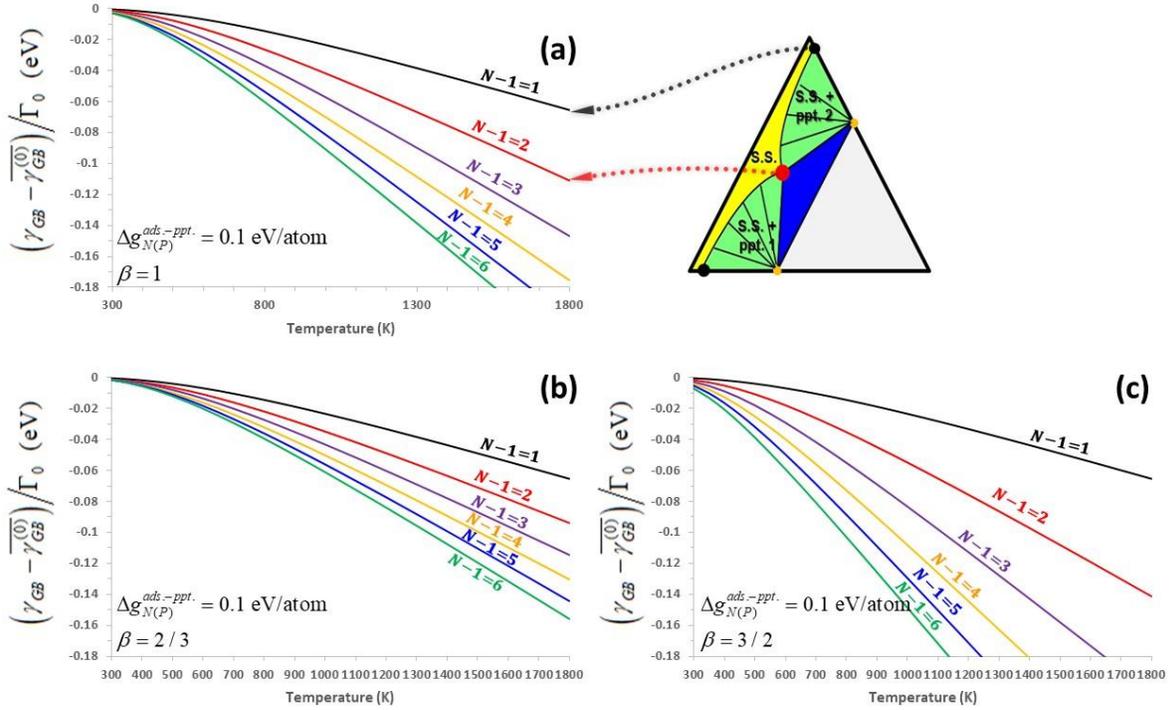

**FIG. 6.** Analysis of Type II HEGBs based on Equation (63) for a hypothetic saturated MPEA (or HEA) with $(N-1)$ principal components plus one segregating minor component. It is assumed, for simplicity, that the primary phase is in equilibrium with $(N-1)$ precipitated secondary phases of binary intermetallic compounds of identical $M_xS_y$ ($\beta = x/y$) stoichiometry and all $(N-1)$ binary subsystems have identical thermodynamic properties. Computed GB energy reduction vs. temperature curves for $\Delta g_{N(P)}^{ads.-ppt.}$ = 0.1 eV/atom and $\beta$ = **(a)** 1, **(b)** 2/3, and **(a)** 3/2, respectively. In all cases, GB energy decreases with increasing temperature (producing a positive effective GB entropy) in saturated MPEAs (HEAs). Furthermore, the effective GB entropy ($-d\gamma_{GB}/dT$) increases with increasing number of components in the MPEA/HEA primary phase. If the $\Gamma_0$ value of the NbMoTaW (110) twist GB in this lattice model is adopted as a reference, $(\gamma_{GB} - \overline{\gamma_{GB}^{(0)}})/\Gamma_0 = -0.18$ eV corresponds to an GB energy reduction of ~1.55 J/m². An isothermal section of ternary phase diagram displaying the saturated compositions for binary ($N-1 = 1$) and ternary ($N-1 = 2$) alloys are also shown. This ternary phase diagram illustrates the increased solid solubility of the segregating element (always set to be Component *N*) with increasing number of components in the primary bulk phase, which results in the increased total GB adsorption. The same mechanism is also expected to provide HEGB effects for $N > 3$.